\documentclass[prd,twocolumn,superscriptaddress,floatfix,amsmath,amssymb,amsfonts,nofootinbib,longbibliography]{revtex4-2}

\usepackage{float} 
\usepackage{scalerel}
\usepackage[normalem]{ulem}
\usepackage[english]{babel}
\usepackage{graphicx}
\usepackage{dcolumn}
\usepackage{bm}
\usepackage{blindtext}
\usepackage{verbatim}
\usepackage{relsize}
\usepackage{mathrsfs}
\usepackage{musicography}
\usepackage{amsmath}
\usepackage{blindtext}
\usepackage{cancel}
\usepackage{physics}
\usepackage{epstopdf}
\usepackage{mathtools}
\usepackage{blindtext}
\usepackage{tensor}
\usepackage{color}
\usepackage[usenames,dvipsnames]{pstricks}
\usepackage{epsfig}
\usepackage{pst-grad} 
\usepackage{pst-plot} 
\usepackage{hyperref}
\usepackage{verbatim}
\usepackage{slashed}
\usepackage{dsfont}
\usepackage{upgreek}

\newcommand{\mf}{\mathsf}

\newcommand{\ii}{\mathrm{i}}

\newcommand{\tc}[1]{\textsc{#1}}

\DeclareMathOperator*{\sumint}{%
\mathchoice%
  {\ooalign{$\displaystyle\sum$\cr\hidewidth$\displaystyle\int$\hidewidth\cr}}
  {\ooalign{\raisebox{.14\height}{\scalebox{.7}{$\textstyle\sum$}}\cr\hidewidth$\textstyle\int$\hidewidth\cr}}
  {\ooalign{\raisebox{.2\height}{\scalebox{.6}{$\scriptstyle\sum$}}\cr$\scriptstyle\int$\cr}}
  {\ooalign{\raisebox{.2\height}{\scalebox{.6}{$\scriptstyle\sum$}}\cr$\scriptstyle\int$\cr}}
}

\allowdisplaybreaks[1] 

\begin{document}

\title{A relativistic QFT description for the interaction of a spin with a magnetic field}

\author{Ruhi Shah}
\email{ruhi.shah@uwaterloo.ca}

\affiliation{Perimeter Institute for Theoretical Physics, Waterloo, Ontario, N2L 2Y5, Canada}
\affiliation{Institute for Quantum Computing, University of Waterloo, Waterloo, Ontario, N2L 3G1, Canada}
\affiliation{Department of Physics and Astronomy, University of Waterloo, Waterloo, Ontario, N2L 3G1, Canada}

\author{Eduardo Mart\'in-Mart\'inez}
\email{emartinmartinez@uwaterloo.ca}

\affiliation{Perimeter Institute for Theoretical Physics, Waterloo, Ontario, N2L 2Y5, Canada}
\affiliation{Department of Applied Mathematics, University of Waterloo, Waterloo, Ontario, N2L 3G1, Canada}
\affiliation{Institute for Quantum Computing, University of Waterloo, Waterloo, Ontario, N2L 3G1, Canada}

\author{T. Rick Perche}
\email{trickperche@perimeterinstitute.ca}

\affiliation{Perimeter Institute for Theoretical Physics, Waterloo, Ontario, N2L 2Y5, Canada}
\affiliation{Institute for Quantum Computing, University of Waterloo, Waterloo, Ontario, N2L 3G1, Canada}
\affiliation{Department of Applied Mathematics, University of Waterloo, Waterloo, Ontario, N2L 3G1, Canada}

\begin{abstract}

We analyze how non-relativistic effective models for the magnetic coupling of a spin to the electromagnetic field (proportional to $\hat{\bm \sigma}\cdot \bm B$) emerge from a full quantum field theoretical description of charged fermionic fields with the quantum electromagnetic field. This allows us to keep track of relativistic corrections to the models commonly used in experimental spin physics. We discuss how this interaction compares to the usual simplified models used in relativistic quantum information.

\end{abstract}

\maketitle

\section{Introduction}

{\color{black}

The fundamental interactions in our Universe are understood to be mediated by quantum fields. These interactions are typically studied through effective theories, which provide valuable insights into the fundamental descriptions within their regimes of validity. While effective descriptions are often sufficient for many physical phenomena, they inherently come with limitations that can obscure important features of the systems they aim to describe. This becomes particularly significant in the context of relativistic quantum information (RQI) protocols that leverage effects that require a more fundamental description. Examples of such protocols that have been extensively studied are entanglement harvesting~\cite{Valentini1991,Reznik1,reznik2,Salton:2014jaa,Pozas-Kerstjens:2015,Pozas2016,HarvestingSuperposed,Henderson2019,bandlimitedHarv2020,ampEntBH2020,mutualInfoBH,threeHarvesting2022,twist2022}, entanglement farming~\cite{Farming}, quantum energy teleportation (QET)~\cite{teleportation,teleportation2014,nichoTeleport,teleportExperiment}, quantum collect calling~\cite{Jonsson2,collectCalling,PRLHyugens2015}.

Even when studying these fundamental protocols, it is common to employ effective (internally) non-relativistic models to describe the localized systems used to implement them. This is because relativistic descriptions of localized bound systems typically arise in non-perturbative regimes of quantum field theory (QFT), often being intractable~\cite{BountStatesQED1951,FieldQuantizationGreiner1996}. A very common approach to describing  local interactions of probes with quantum fields is the Unruh-DeWitt (UDW) detector~\cite{Unruh1976,DeWitt}. The UDW model simplifies the interaction of a localized system with a quantum field by considering a quantum system (very commonly a qubit) interacting with a quantum scalar field. This model captures some of the essential aspects of the local interaction in QFT, for example light-matter interactions~\cite{eduardoOld2013,Nicho1,richard}. However, the model also has its drawbacks. For instance, it is well-known that non-relativistic probes can have issues with relativistic covariance~\cite{us,us2} and causality~\cite{EduCusality2015,PipoFTL,mariaPipoNew} outside of their regimes of validity. Moreover, the commonly employed two-level UDW detector is the ``wildcard'' of QFT probes, versatile, but naturally failing to incorporate aspects of specific physical processes~\cite{Pozas2016,richard}. For practical implementations of RQI protocols, one then has to adapt the model to match the specific experimental setup at hand.

Examples of systems whose fundamental description naturally reduces to a qubit can be found in the spin degrees of freedom of many fermions such as electrons, making them suitable candidates for implementation of relativistic quantum information protocols. 
Moreover, the electron spin can be measured and controlled with high precision for quantum information processing and quantum sensing on a variety of experimental platforms, including electron spin resonance~\cite{esrReviewDavid, esrspincavity} and nitrogen-vacancy centers in diamond~\cite{NV-centers-electron-spin-control}.

However, when describing spin interactions it is usual to employ non-relativistic approximations, such as the Zeeman interaction and J-coupling~\cite{sakurai}, often treating the electromagnetic potential as a classical background or replacing it by a direct coupling between spins. These not only do not consider relativistic aspects of the interaction, they omit the field dynamics altogether. These approximations, while practical, neglect important aspects of QFT that become crucial when implementing relativistic quantum information protocols that rely on the quantum degrees of freedom of the electromagnetic field.

In this paper, we will study the interaction of a quantum spin with an external electromagnetic field starting from a fully quantum field theoretic perspective. This approach gives control over non-relativistic approximations that reduce quantum electrodynamics to effective simpler models, such as the Zeeman Hamiltonian. Carefully taking the steps that reduce the fundamental description of spins to these effective models clearly specifies their regime of validity and gives correction terms that arise from the complete quantum field theoretical approach. 

By incorporating relativistic effects into the modelling of a localized spin, we obtain a relativistic quantum field theory description of a localized quantum probe of the magnetic field. This approach serves as a natural connection between fundamental descriptions of measurement apparatuses in QFT~\cite{FewsterVerch,fewster2} and the more operational perspective based on effective models~\cite{chicken,hectorChicken}. We will compare the relativistic decription of the spin-magentic coupling with the ubiquitous two-level UDW detector. Perhaps surprisingly, we find that this magnetic field spin detector model is in many ways simpler than the UDW model due to its symmetries. This paves the way for studies and applications of relativistic quantum information protocols utilizing the electron spin as a quantum probe of the electromagnetic field.

}

This manuscript is organized as follows. In Section~\ref{sec:QED-Atom} we review the description of the Dirac atom, describing the electron as a localized mode excitation of a relativistic fermionic field. In Section~\ref{sec:C2reduction} we show how the spin of an $s$-orbital electron in a hydrogen-like atom can be described as an effective two-level system and discuss the consequences of the reduction of the QFT description of the electron to a spin degree of freedom. In Section~\ref{sec:coupleExtMag} we show how the typical Zeeman interaction with an external magnetic field emerges form the fundamental description. Section~\ref{sec:quantumProbe} is devoted to exploring finite time interactions of a qubit with an external quantum magnetic field and comparing this model with the commonly employed two-level Unruh DeWitt detector.
The conclusions of our work can be found in Section~\ref{sec:Conclusions}.

\section{Second-Quantized Dirac Hydrogen Atom}\label{sec:QED-Atom}

In this section we review the description of a second-quantized Dirac atom, which models the electron as a quantum field under the influence of a classical Coulomb potential sourced by the nucleus. The equations of motion for the Dirac field then admit a countable number of bound solutions, which are effectively localized around the central potential. This way of modelling Hydrogen-like atoms has been thoroughly studied, and a detailed description can be found in standard textbooks~\cite{RQMgreiner, antonRQM}. 

In 3+1 Minkowski spacetime, a Dirac spinor $\psi^a(\mathsf{x})$, for $a \in \{1,\,2,\,3,\,4\}$ can be represented as a four-component complex field with a 
 representation of $\text{SL}_2(\mathbb{C})$ (the universal cover of the Lorentz group SO(1,3)). A Lorentz transformation $\Lambda$ acts on spinors according to
\begin{equation}
    \psi'{}^a(\mf x) = S[\Lambda]^a_{\,\,b}\psi^b(\Lambda^{-1}\mf x) \,,
\end{equation}
where $S[\Lambda]$ is given by $S[\Lambda] = \exp(\frac{1}{2}\omega_{\mu\nu}S^{\mu\nu})$ and $S^{\mu\nu}$ are the generators of the $\text{SL}_2(\mathbb{C})$ action. Notice that for any values of $\mu,\nu\in\{0,1,2,3\}$, each of these generators is an operator in spinor space. They can be conveniently expressed in terms of the so-called gamma matrices, defined by the Clifford algebra relation
\begin{equation}
    \gamma^\mu\gamma^\nu + \gamma^\nu\gamma^\mu = -2\eta^{\mu\nu}\,,
\end{equation}
where $\eta_{\mu\nu} = \text{diag}(-1,1,1,1)$ is the Minkowski metric in diagonal form. The generators $S^{\mu\nu}$ can then be written as \mbox{$S^{\mu\nu} = \frac{1}{4}[\gamma^\mu, \gamma^\nu]$}.

The dynamics of a free spinor field  $\psi(\mf x)$ of mass $m_e$ is given by the Dirac equation, which, in inertial coordinates $(t,\bm x)$, takes the form
\begin{equation}
    (\ii\slashed{\partial} - m_e)\psi(\mf x) = 0 \,.
\end{equation}
Here, we use the Feynman slashed notation $\slashed{\partial} = \gamma^\mu \partial_\mu$.
In the Dirac representation the $\gamma$-matrices are,
\begin{equation}\label{gammamatrices} 
\gamma^0=\begin{pmatrix}
\mathds{1}_2 & \\
& -\mathds{1}_2
\end{pmatrix}, \quad 
\gamma^i=\begin{pmatrix}
  & \sigma^i \\
-\sigma^i & 
\end{pmatrix} \, ,
\end{equation}
where $\bm \sigma = (\sigma_x,\sigma_y,\sigma_z)$ denote the Pauli matrices and $\openone_2$ is the $2\times2$ identity matrix.

To couple a Dirac spinor to electromagnetism, we consider the U(1) gauge transformation associated with the charge $- q$,
\begin{equation}
    \psi(\mf x) \mapsto e^{\ii q\alpha(\mf x)} \psi(\mf x),
\end{equation}
which is generated by the electromagnetic four-potential $A = A_\mu \text{d}x^\mu$. 
The corresponding Lagrangian density for a Dirac spinor $\psi$ minimally coupled to an electromagnetic four-potential $A_\mu$ is,
\begin{equation}\label{QED-lagrangian-density}
    \mathscr{L} = \bar{\psi}(\ii\gamma^\mu D_\mu - m_e)\psi  - \frac{1}{4} F_{\mu\nu}F^{\mu\nu}\, .
\end{equation}
Here $\bar{\psi} \equiv\psi^\dagger \gamma^0$, $D_\mu \equiv\partial_\mu - \ii q A_\mu$ is the covariant derivative with respect to the U(1) gauge transformation,  and \mbox{$F_{\mu\nu} = \partial_\mu A_\nu - \partial_\nu A_\mu$} is the electromagnetic field strength tensor.


To describe the electron field in an atom, we model the nucleus as a non-dynamical static point charge with four-current density $j^\mu(\mf x)$. The full Lagrangian for the theory then becomes
\begin{equation}\label{QED-lagrangian-density2}
    \mathscr{L} = \bar{\psi}(\ii\gamma^\mu D_\mu - m_e)\psi  - \frac{1}{4} F_{\mu\nu}F^{\mu\nu} - j^\mu A_\mu\, .
\end{equation}
The corresponding equations of motion are
\begin{align}
    \partial^\mu F_{\mu\nu} = j_\nu -q \bar{\psi}\gamma_\nu \psi,\\
    (\ii \slashed{\partial} - m_e)\psi = -q\slashed{A} \psi.
\end{align}
The solutions to the equations of motion for the four-potential $A_\mu$ can be written as a sum of the solutions to the free part $A_\mu^{(\text{free})}$, and the part sourced by the nucleus $A^{(\text{atom})}_\mu$,
\begin{align}\label{eq:Adecomp}
    A_\mu = A_\mu^{(\text{free})} + A^{(\text{atom})}_\mu \,, 
\end{align}
which satisfy
\begin{align}
    \partial^\mu F^{(\text{atom})}_{\mu\nu} = j_\nu,\\
    \partial^\mu F^{(\text{free})}_{\mu\nu} = -q \bar{\psi}\gamma_\nu \psi. 
\end{align}

We fix an inertial frame $(t,\bm x)$, comoving with the nucleus, and prescribe
\begin{equation}\label{eq:nucleusJmu}
    j^\mu(\mf x) = Q u^\mu \delta^{(3)}(\bm x ),
\end{equation}
where $u^\mu = (1,0,0,0)$ is the four-velocity of the nucleus and $Q = {qZ}$ is its charge. The pointlike charge then sources a Coulomb potential $A^{(\text{atom})}_\mu(r)$, where $r = |\bm x|$ is the radial coordinate centered at the nucleus, $r = \sqrt{x^i x_i}$. In the Coulomb gauge, the potential is given by
\begin{equation}\label{atom-potential}
    A_{\mu}^{(\text{atom})}(t, \bm{x}) = 
        -  \frac{Q}{r} u_\mu.
\end{equation}

The electron orbitals can be found by solving the equations of motion for the $\psi(\mf x)$ field only considering the Coulomb potential sourced by the nucleus: 
\begin{equation}\label{Dirac-eqn-atom}
    0=(\ii\slashed{\partial} +  q\slashed{A}^{(\text{atom})}(\mathsf{x}) - m_e)\psi(\mathsf{x}) \, .
\end{equation}
To solve the equation and obtain the basis of solutions for the electron, we can split the time and space components of the equation in this inertial frame, 
\begin{equation}\label{Dirac-eqn-atom-split}
    \ii\partial_0\psi(\mathsf{x}) = (-\ii\gamma^0\gamma^i\partial_i - \gamma^0m_e - qA_0(r))\psi(\mathsf{x}) \, .
\end{equation}
A general solution to this equation can be found by looking for static solutions which are eigenfunctions of the Hamiltonian
\begin{equation}\label{H-Dirac-atom}
    \mathsf{H}_{\text{atom}} = (-\ii\gamma^0\gamma^i\partial_i + \gamma^0m_e - qA_0(r)) \, .
\end{equation}
That is, we look for solutions to the following time-independent Dirac equation,
\begin{equation}\label{TIDE-atom}
    \mathsf{H}_{\text{atom}}\psi(\bm{x}) = E\psi(\bm{x}) \, .
\end{equation}
The Hamiltonian $\mathsf{H}_\text{atom}$ is analogous to the familiar Hamiltonian of a Schrodinger hydrogen atom. In this case, the Hamiltonian acts on a spinor-valued field, and its eigenfunctions correspond to the \emph{classical} static solutions. 

The solutions to this eigenvalue problem can be classified in terms of eigenvalues of a set of operators that commute with $\mathsf{H}_\text{atom}$. The following operators are relevant to this problem: 
\begin{align}\label{Dirac-atom-operators}
    \bm{\mf{J}} &= \bm{\mf{L}} + {\bm \Upsigma}\,, \quad\quad {\bm{\mf{L}}} = - \ii \bm{r} \times \bm \nabla \, , \quad\quad {\bm \Upsigma} = \begin{pmatrix}
        \bm \sigma & 0 \\
        0 & \bm \sigma
    \end{pmatrix} \,,
\end{align}
corresponding to total angular momentum ($\bm{\mf{J}}$), orbital angular momentum ($\bm{\mf{L}}$), and spin ($\bm{\Upsigma}$). In addition, we define the parity operator,
\begin{equation}\label{Dirac-parity-operator}
    \mf{P}\psi(t, \bm{x}) = \gamma^0\psi(t, -\bm{x}) \, .
\end{equation} 
Both the parity operator and the total angular momentum operator commute with the Hamiltonian $\mf{H}_{\text{atom}}$, as well as $\mf{J}_z$,  
\begin{equation}\label{Dirac-atom-commute}
    [\bm{\mf{J}}, \mf{H}_{\text{atom}}] = [\mf{J}_z, \mf{H}_{\text{atom}}] = [\mf{P}, \mf{H}_{\text{atom}}] = 0 \,.
\end{equation}
Together with $\mf{H}_\text{atom}$, their eigenvalues are enough to label all bound stationary solutions of Dirac's hydrogen atom. Notice that
although the total angular momentum is conserved, the orbital angular momentum ($\bm{\mf{L}}$) and spin ($\bm \Upsigma$) are not. The solutions of the equation of motion can therefore be labelled by four quantum numbers, $n$, $j$, $m$, and $p$, defined by
\begin{align}
    \mf{J}^2\psi_{njmp} &= j(j+1)\psi_{njmp} \, , \\
    \mf{J}_z\psi_{njmp} &= m\psi_{njmp} \, ,  \\
    \quad\quad\quad\quad\quad\quad\quad\mf{P}\psi_{njmp} &= p\psi_{njmp} \, , \\ \quad\quad\quad\quad\mf{H}_{\text{atom}}\psi_{njmp} &= E_{nj}\psi_{njmp}\label{TIDE},
\end{align}
where 
\begin{equation}\label{Energies-dirac-atom}
    E_{nj} =\frac{m_e}{\sqrt{1+\frac{(Z \alpha)^2}{\left(n-j-\frac{1}{2}+\sqrt{\left(j+\frac{1}{2}\right)^2-(Z \alpha)^2}\right)^2}}}
\end{equation}
and $\alpha$ is the fine structure constant. The quantum numbers $n$, $j$, $m$, and $p$ take the discrete values
\begin{align}\label{Dirac-atom-eigenvalues}
    n &= 1,\,2\,,...\,,\\
    j &= \frac{1}{2},\,...\,,\,n-\frac{1}{2}\,, \\
    m &= -j, -(j-1), ..., j-1, j \, , \\
    p &= \begin{cases}
        +1, \text{ if } j = n - 1/2,\\
        \pm 1, \text{ if } j \neq n - 1/2        
    \end{cases} .
\end{align}
Notice that unlike the case of Schrodinger's hydrogen atom, in Dirac's atom the energy levels depend on the total orbital quantum number $j$.

As with any time-independent spherically symmetric external potential $A_0(r)$, the four-component Dirac eigenfunctions $\psi_{njmp}$ can be split into two two-component bispinors,
\begin{equation}\label{psi-sol-atom}
    {\color{black}\psi_{njmp}(\bm x) = \begin{pmatrix}
        g_{nj}(r)\Omega_{jml}(\theta,\phi)\\
        f_{nj}(r)\Omega_{jml'}(\theta,\phi)
    \end{pmatrix} \, .}
\end{equation}
Here the functions $g_{nj}(r)$ and $f_{nj}(r)$ define the effective localization lengthscale of the modes and $\Omega_{jml}$ are the spinor spherical harmonics~\cite{RQMgreiner, spinorsphericalharmonics}, where $l$ labels orbital angular momentum according to $\bm{\mf{L}}^2\Omega_{jml} = l(l+1)\Omega_{jml}$. In Eq.~\eqref{psi-sol-atom} $l$ and $l'$ can take the values of $j \pm \tfrac{1}{2}$. $l$ and $l'$ are related by the parity eigenvalue $p = \pm 1$ through $l' - l =  p$.  



The operator $\mf{H}_\text{atom}$ also possesses a continuous spectrum for $E\geq m_e$, representing scattering states. These can be labelled by a continuous parameter $k\geq 0$ and the quantum numbers $j,m,p$. The corresponding eigenfunctions can be classified by the sign of the eigenvalues of the operator $\ii \partial_t$. We will notate them as $u_{kjmp}(\bm x)$ for positive frequencies (electron states) and $v_{kjmp}(\bm x)$ for negative frequencies (positron states). The explicit expressions for the scattering solutions can be found in~\cite{RQMgreiner}.

We can then write a general solution of Eq.~\eqref{Dirac-eqn-atom} as a mode expansion in terms of the solutions $\psi_{\bm N}(\bm x)$, as well as a continuous set of unbound solutions $u_{\bm k}(\bm x)$ and $v_{\bm k}(\bm x)$, where we use the multi-indices $\bm N = (n,j,m,p)$, $\bm k = (k,j,m,p)$ and $k$ is a continuous parameter. Overall, a general solution of the equation of motion can be written in terms of coefficients $b_{\bm N}$, {${b}_{\bm k}$} and {${c}_{\bm k}$} as
\begin{align}\label{Dirac-solution}
    \psi(\mf x) = &\sum_{\bm N} b_{\bm N} e^{-\ii E_{\bm N}t}\psi_{\bm N}(\bm{x})
    \\& + \sumint_{\bm k} \left( b_{\bm k} e^{- \ii E_{\bm k} t} u_{\bm k}(\bm x) + c_{\bm k}^* e^{\ii E_{\bm k} t} v_{\bm k}(\bm x)\right).\nonumber
\end{align}

Now that we have a full solution to (\ref{Dirac-eqn-atom}), we can second quantize the coefficients in this solution to build a Fock space for the electron states. That is, we promote the coefficients in (\ref{Dirac-solution}) to operators, \mbox{$b_{\bm N}\rightarrow\hat{b}_{\bm N}$}, \mbox{$b_{\bm k}\rightarrow\hat{b}_{\bm k}$}, \mbox{$c_{\bm k}\rightarrow\hat{c}_{\bm k}$}. The operators $\hat{b}_{\bm N},\, \hat{b}_{\bm k},\, \hat{c}_{\bm k}$ are the annihilation operators and $\hat{b}^\dagger_{\bm N},\, \hat{b}^\dagger_{\bm k},\, \hat{c}^\dagger_{\bm k}$ are the corresponding creation operators. These operators are defined by the canonical anti-commutation\footnote{The fermionic field for the electron is a spin-$\frac{1}{2}$ field, so that the {spin-statistics theorem}~\cite{spin-statistics1958,ArakiSpinStatistics1961} imposes anti-commutation relations for these operators. The anti-commutation relations ensure that resulting quantum field theory has a Hamiltonian bounded from below.} relations,
\begin{align}
    \{\hat{b}_{\bm N}, \hat{b}^\dagger_{\bm N'}\} &= \delta_{\bm{NN}'},\label{eq:commNN}\\
    \{\hat{b}_{\bm k}, \hat{b}^\dagger_{\bm k'}\} &= \delta^{(3)}(\bm k - \bm k'),\\
    \{\hat{c}_{\bm k}, \hat{c}^\dagger_{\bm k'}\} &= \delta^{(3)}(\bm k - \bm k').
\end{align} 
Notice that the bound state solutions (characterized by the discrete quantum numbers $\bm N$) satisfy Kronecker-delta anticommutation relations, while the operators associated to the scattering modes anticommute to Dirac-deltas. Eq.~\eqref{eq:commNN} reflects the fact that the discrete modes $\psi_{\bm N}(\bm x)$ are orthonormal according to the Dirac inner product
\begin{equation}
    (\psi_{\bm N}, \psi_{\bm N'}) = \int \dd^3 \bm x \bar{\psi}_{\bm N}(\bm x) \gamma^0 \psi_{\bm N'}(\bm x)  = \delta_{\bm N \bm N'}.
\end{equation}
The creation and annihilation operators define a vacuum $\ket{0}$ by the relations $\hat{b}_{\bm N} \ket{0} = \hat{b}_{\bm k} \ket{0} = \hat{c}_{\bm k} \ket{0} = 0$ for all $\bm N$ and $\bm k$. The electron quantum field can then be written as
\begin{align}\label{psi-general}
    \hat{\psi}(\mf x) = &\sum_{\bm N} \hat{b}_{\bm N} e^{-\ii E_{\bm N}t}\psi_{\bm N}(\bm{x})
    \\& + \sumint_{\bm k} \left( \hat{b}_{\bm k} e^{- \ii E_{\bm k} t} u_{\bm k}(\bm x) + \hat{c}^\dagger_{\bm k} e^{\ii E_{\bm k} t} v_{\bm k}(\bm x)\right).\nonumber
\end{align}
In the context of this quantum field theory, the Hamiltonian density that prescribes the dynamics of the electron field under the influence of the Coulomb potential is
\begin{equation}
    \hat{\mathcal{H}}_{\text{atom}}(\mf x) = \hat{\bar{\psi}}(\mf x)(-\ii\gamma^i\partial_i + m_e + q \slashed{A}^{\text(atom)})\hat{\psi}(\mf{x}),
\end{equation}
not to be confused with the operator $\mf{H}_\text{atom}$, which is a differential operator that acts in classical solutions of the equation of motion. The corresponding Hamiltonian of the quantum theory for the fermionic field under the influence of the Coulomb potential can be found by regularizing\footnote{The fact that the Hamiltonian density $\hat{\mathcal{H}}_\text{atom}(\mf x)$ is quadratic in the field operator $\hat{\psi}(\mf x)$ implies that its expected value in any field state is, in principle, divergent, requiring regularization. One common method of regularizing observables of this type is to subtract the expected value of the operator at a reference state.} and integrating the Hamiltonian density along a spatial slice $t = \text{const.}$We regularize the Hamiltonian by subtracting its expected value in the vacuum $\ket{0}$ defined by the expansion of Eq.~\eqref{psi-general}:
\begin{align}\label{H-atom-quantized}
    &\hat{H}_{\text{atom}} \coloneqq \int \dd^3 \bm x\, \Big(\hat{\mathcal{H}}_{\text{atom}}(\mf x) - \bra{0}\hat{\mathcal{H}}_{\text{atom}}(\mf x)\ket{0} \Big) \nonumber \\
    &= \sum_{\bm N} E_{\bm N}\hat{b}^\dagger_{\bm N}\hat{b}_{\bm N} + \sumint_{\bm k} E_{\bm{k}} (\hat{b}_{\bm k}^\dagger\hat{b}_{\bm k} + \hat{c}_{\bm k}^\dagger\hat{c}_{\bm k}).
\end{align}
The eigenstates of the Hamiltonian can be constructed by repeated applications of the creation operators on the vacuum state $\ket{0}$, and these eigenstates span the field's Fock space, $\mathcal{F}_\psi$. The one-particle states of the theory are
\begin{align}
    \ket{\bm N} &= \hat{b}_{\bm N}^\dagger \ket{0},\\
    \ket{\bm k,+} &= \hat{b}_{\bm k}^\dagger \ket{0},\\
    \ket{\bm k,-} &= \hat{c}_{\bm k}^\dagger \ket{0}.
\end{align}
Notice that the states $\ket{\bm k,\pm}$ are not normalizable, as they correspond to the continuous spectrum of $\hat{{H}}_\text{atom}$. Therefore the scattering states are not physical, although they form a useful basis for the study of scattering processes. On the other hand, the bound states $\ket{\bm N}$ are localized physical states, satisfying
\begin{equation}\label{eq:wearenotparticlephysicists}
    \braket{\bm N}{\bm N'} = \delta_{\bm N\bm N'}.
\end{equation}
These states correspond to electrons bound to the hydrogen atom with quantum numbers $\bm N = (n,j,m,p)$.

While Eq.~\eqref{psi-general} allows one to describe an electron in terms of a fully featured quantum field, it does not describe aspects of the electron-proton interaction. For instance, the hyperfine splitting in the electron energy levels comes from the interaction of the spins of the proton and electron through the magnetic field, which is not present in the description above\footnote{It is possible to effectively implement the hyperfine splitting by considering an additional $\mathbb{C}^2$ quantum degree of freedom corresponding to the proton spin and introducing an appropriate dipole coupling with the field $\psi(\mf x)$~\cite{RQMgreiner}.}. 

The lack of the description for the degrees of freedom of the proton in this model also implies that the bound states in the hydrogen atom are electron states. This is unlike the Schr\"odinger atom description, where one defines both center of mass and internal degrees of freedom for the system, mixing the electron and proton wavefunctions~\cite{richard}. That is, the bound states of a Schr\"odinger atom do not correspond to the electron degrees of freedom per s\'e, but to a combination of the electron and proton systems. In order to obtain this feature in a QFT description, one would need to consider both the electron and proton as fermionic fields, which would require more sophisticated bi-spinor techniques~\cite{bispinorQED}, and falls beyond the scope of this work.

Finally, notice that although the electron field is fully relativistic, this model clearly privileges the reference frame of the nucleus. This implies that the solutions of the electronic field are not Poincar\'e invariant, instead only being invariant under time translations in the direction of the nucleus four-velocity $u^\mu$ and rotations around the origin in the nucleus' rest space. However, the lack of Poincar\'e covariance is to be expected: no localized system can be invariant under translations or arbitrary boosts. On the other hand, symmetry under arbitrary transformations generated by the Poincar\'e group can be restored by also applying these transformations to the nucleus ($j^\mu(\mf x)$) and Coulomb field. In other words, the field representation of Eq.~\eqref{psi-general} is valid the any inertial frame where the nucleus current density takes the shape of Eq.~\eqref{eq:nucleusJmu}.

\section{Reducing a relativistic atom to $\mathbb{C}^2$}\label{sec:C2reduction}

In this section we reduce the quantum field description of the Hydrogen atom presented above to an effective two-level system. To do this, we restrict the quantum numbers of the fermionic field~\eqref{Dirac-eqn-atom} to two degrees of freedom corresponding to the one-particle sector of the $s$ orbital of an atom. 

In the usual Schr\"odinger description of a Hydrogen atom the $s$ orbitals are defined by the vanishing of the quantum number associated with orbital angular momentum ($l=0$). However, in the Dirac description, the orbital angular momentum does not define a quantum number, as it does not commute with the atom Hamiltonian. Instead, in this description, the $s$ orbitals correspond to the quantum numbers $j=  1/2$ and $p = +1$~\footnote{To compare the orbitals defined by a Dirac hydrogen-like atom with the ones in the Schr\"odinger description, it is necessary to employ non-relativistic approximations. These are usually done through a Foldy-Wouthuysen~\cite{FoldyWou,RQMgreiner} transformation. In this framework the top two components of the Dirac spinor, which have well-defined orbital angular momentum, are dominant. This implies that, in the non-relativistic approximation, the Dirac spinor describing a state with $j=\frac{1}{2}$ and $l=0$ has positive parity, since the dominant top component has $l=0$ and the bottom component has $l'=1$, yielding $p = l'-l = +1$.}. The two degrees of freedom in a given $s$ orbital are encoded in the magnetic quantum number $m$, which can take the values $m = \pm\frac{1}{2}$. For instance, the $1s$ orbital corresponds to the to the subspace defined by the quantum numbers $n=1$ and $j=1/2$, which imply $p=+1$.

To reduce the quantum field description to a specific $s$ orbital, we fix the quantum number $n = n_0$. For convenience, we introduce the following notation for the operators and states associated with the quantum numbers $(n,j,m,p) =(n_0,\frac{1}{2},\pm\tfrac{1}{2},+1)$:
\begin{equation}
    \hat{b}_\uparrow \coloneqq \hat{b}_{n_0,\frac{1}{2},\frac{1}{2},+1},\quad\quad
    \hat{b}_\downarrow \coloneqq \hat{b}_{n_0,\frac{1}{2},-\frac{1}{2},+1}.
\end{equation}
\vspace{-10mm}

\begin{equation}
\begin{aligned}\label{uparrowdownarrowstates}
    \ket{\uparrow}\, &\coloneqq  \ket{n_0,\tfrac{1}{2},\tfrac{1}{2},+1} = \hat{b}^\dagger_{\uparrow}|0\rangle ,\\
    \ket{\downarrow}\, &\coloneqq  \ket{n_0,\tfrac{1}{2},-\tfrac{1}{2},+1} = \hat{b}^\dagger_{\downarrow}|0\rangle  .
    \end{aligned}
\end{equation}
The fact that the modes of the field $\hat{\psi}(\mf x)$ are discrete implies that the states $\ket{\uparrow}$ and $\ket{\downarrow}$ are normalized (see Eq.~\eqref{eq:wearenotparticlephysicists}). $\ket{\uparrow}$ and $\ket{\downarrow}$ form an orthonormal basis for the two dimensional subspace that they span, $\mathcal{H}_\text{s}\cong \mathbb{C}^2$. Within this subspace, we define ladder operators $\hat{\sigma}_+$ and $\hat{\sigma}_-$ as
\begin{align}
    \hat{\sigma}_+ &\coloneqq \ket{\uparrow}\!\bra{\downarrow}\,, \quad \hat{\sigma}_- \coloneqq \ket{\downarrow}\!\bra{\uparrow} \,,
\end{align}
and the operators 
\begin{align}
    \hat{\sigma}_x &\coloneqq \hat{\sigma}_+ + \hat{\sigma}_- \,,\\
    \hat{\sigma}_y &\coloneqq -\ii(\hat{\sigma}_+ - \hat{\sigma}_-), \,\\
    \hat{\sigma}_z &\coloneqq \hat{\sigma}_+ \hat{\sigma}_- - \hat{\sigma}_-\hat{\sigma}_+,
\end{align}
which satisfy the $\mathfrak{su}(2)$ algebra commutation relations
\begin{equation}
    [\hat{\sigma}_i, \hat{\sigma}_j] = 2\ii\epsilon_{ij}^{\,\,\,\,k}\,\hat{\sigma}_{k} \, ,
\end{equation}
so that they act as the Pauli operators in $\mathcal{H}_\text{s}$ and their matrix representation in the basis $\{\ket{\uparrow},\ket{\downarrow}\}$ takes the form
\begin{equation}
    \hat{\sigma}_x = \begin{pmatrix}
        0&&1\\
        1&&0
    \end{pmatrix} ,\, \hat{\sigma}_y =\begin{pmatrix}
        0&&-\ii\\
        \ii&&0
    \end{pmatrix},\,
    \hat{\sigma}_z  = \begin{pmatrix}
        1&&0\\
        0&&-1
    \end{pmatrix}\,.
\end{equation}

We define the projector into the subspace $\mathcal{H}_s$,
\begin{equation}
    \hat{P}_{\text{s}} = \ket{\uparrow}\!\bra{\uparrow}+\ket{\downarrow}\!\bra{\downarrow}. \, 
\end{equation}
The projector $\hat{P}_\text{s}$ can be used to reduce operators acting on the Fock space $\mathcal{F}_\psi$ to $\mathcal{H}_\text{s}\cong \mathbb{C}^2$. For instance, consider an operator acting on the Fock space $\mathcal{F}_\psi$ of the form, 
\begin{equation}\label{generaloperatorM}
    \hat{M}(\mf x) = \hat{\bar{\psi}}(\mf{x})O(\mf{x})\hat{\psi}(\mf{x})\,,
\end{equation}
where $O(\mf{x})$ is an operator that acts on (the classical) spinor space. Let \mbox{${M_{\bm{N}\bm{N}'}}(\mf{x}) \coloneqq \bar{\psi}_{\bm{N}}(\mf{x}) O(\mf{x})\psi_{\bm{N}'}(\mf{x})$}, where $\bm{N}, \, \bm{N}'$ are labels for the  quantum numbers $(n,\,j,\, m,\,p)$ introduced in the previous section. The operator $\hat{M}(\mf x)$ can be generally written as
\begin{align}\label{Moperator}
    \hat{M}(\mf x) &= \sum_{\bm{N},\,\bm{N}'} M_{\bm{N}\bm{N}'}(\mf{x})\hat{b}^\dagger_{\bm{N}}\hat{b}_{\bm{N}'}\,.
\end{align}
The action of the projector $\hat{P}_\text{s}$ on the operators $\hat{b}^\dagger_{\bm{N}}\hat{b}_{\bm{N}'}$ in this expansion can be written in the basis $\{\ket{\uparrow},\ket{\downarrow}\}$ as 
\begin{equation}
    \hat{P}_\text{s}\hat{b}_{\bm N}^\dagger \hat{b}_{\bm N'} \hat{P}_\text{s} = \sum_{m \in \{\uparrow, \downarrow\}} \sum_{ m' \in \{\uparrow, \downarrow\}} \delta_{m \bm{N}} \delta_{m' \bm{N}'}\ket{m}\!\bra{m'}\, ,
\end{equation}
so that applying the projector $\hat{P}_\text{s}$ on Eq.~\eqref{generaloperatorM} yields
\begin{align}
    \hat{M}_{\text{s}}(\mf x) \coloneqq \hat{P}_\text{s}\hat{M}(\mf x)\hat{P}_\text{s} &= M_{\uparrow\uparrow}(\mf{x})\ket{\uparrow}\!\bra{\uparrow} + M_{\uparrow\downarrow}(\mf{x})\ket{\uparrow}\!\bra{\downarrow} \nonumber\\*
    &+ M_{\downarrow\uparrow}(\mf{x})\ket{\downarrow}\!\bra{\uparrow} + M_{\downarrow\downarrow}(\mf{x})\ket{\downarrow}\!\bra{\downarrow}\,,
\end{align}
with matrix representation
\begin{equation}
    \hat{M}_\text{s}(\mf x) = \begin{pmatrix}
        M_{\uparrow\uparrow}(\mf{x})&&M_{\uparrow\downarrow}(\mf{x})\\
        M_{\downarrow\uparrow}(\mf{x})&&M_{\downarrow\downarrow}(\mf{x})
    \end{pmatrix}\, .
\end{equation}
From this point on, we use the subindex s to indicate projection to $\mathcal{H}_\text{s}$.

As an example, consider the second-quantized Hamiltonian (\ref{H-atom-quantized}). We can build its projection on the subspace $\mathcal{H}_\text{s}$, \mbox{$\hat{H}_{\text{atom,}\text{s}}=\hat{P}_\text{s} \hat{H}_{\text{atom}}\hat{P}_\text{s}$}, given by
\begin{equation}\label{H-atom-lzero}
    \hat{H}_{\text{atom,}\text{s}} = \sum_{m = \pm \frac{1}{2}} E_{m}\hat{b}_{m}^\dagger\hat{b}_m = E_{\uparrow}\hat{b}_{\uparrow}^\dagger\hat{b}_\uparrow + E_{\downarrow}\hat{b}_{\downarrow}^\dagger\hat{b}_\downarrow \,.
\end{equation}
Notice that since we are not considering any hyperfine interactions or any external electromagnetic fields, spins up and down have the same energy, $E_{\uparrow} = E_\downarrow$, yielding a $\hat{H}_{\text{atom,}\text{s}}$ proportional to the identity in $\mathcal{H}_\text{s}$. Defining $E_0 \coloneqq E_{\uparrow} = E_\downarrow$, a matrix representation for the atom Hamiltonian in this subspace is then 
\begin{equation}
    \hat{H}_{\text{atom,}\text{s}} = \begin{pmatrix}
        E_0&&0\\0&&E_0
    \end{pmatrix} = E_0\hat{\mathds{1}} \, .
\end{equation}

Reducing the electron field theory to the two-dimensional $\mathcal{H}_\text{s}$ subspace considerably simplifies its dynamics by disregarding particle creation effects and transitions to higher energy modes. Although the states $\ket{\uparrow}$ and $\ket{\downarrow}$ still represent mode excitations of a relativistic field, it is important to notice that the projector $\hat{P}_\text{s}$ is non-local in spacetime. This is because projectors of the form $\ket{\uparrow}\!\bra{\uparrow}$, $\ket{\downarrow}\!\bra{\downarrow}$ are intrinsically non-local, leading to covariance and causality violations which are controlled by the size of the localization of the field modes. This phenomenon has also been noticed in~\cite{QFTPD}, and it is a general fact  that restricting a quantum field theory theory to modes of energies below a certain cutoff in a given frame introduces causality violations~\cite{BurgessEffectiveFieldTheory}. Nevertheless, the reduction of the field theory to the subspace $\mathcal{H}_\text{s}$ is justified whenever one considers processes that effectively take place at low energy and affecting only the $s$ orbital, such as the interaction of a spin in an $s$ orbital with an external electromagnetic field that does not produce mode excitations. It is in this regime where one obtains the leading order relativistic corrections to the hydrogen-like atom~\cite{Dirac-electron-theory,RQMgreiner}.

\section{An Electron Coupled to an External Electromagnetic Field}\label{sec:coupleExtMag}

As an example of the application of the reduction method introduced in the previous section, we derive the Zeeman Hamiltonian 
\begin{equation}
    \hat{H}_I = - \gamma\, \hat{\bm \sigma}\cdot \bm B,\label{eq:ZeemanBase}
\end{equation} 
for the interaction of a spin with an external electromagnetic field starting from the quantum field theoretic description of the electron. The coupling of a fermionic field with electromagnetism is encoded in the Lagrangian density of Eq.~\eqref{QED-lagrangian-density}, which contains the interaction term between $\psi(\mf x)$ and an external electromagnetic field $A^{\text{(ext)}}_\mu(\mf x)$ (see Eq.~\eqref{eq:Adecomp}). For simplicity, for the remainder of the manuscript, we will denote the external field simply by $A_\mu(\mf x)$. The interaction Lagrangian density can then be written as
\begin{equation}
    \mathscr{L}_I(\mf x) = q\bar{\psi}(\mf x) \slashed{A}(\mf x)\psi(\mf x).
\end{equation}
In the quantum field theory description of $\hat{\psi}(\mf x)$, we can then write the associated Hamiltonian density as
\begin{equation}\label{h-int-density}
    \hat{h}_I(\mf x) = - 
 q\hat{\bar{\psi}}(\mf x) \slashed{A}(\mf x)\hat{\psi}(\mf x).
\end{equation}
The interaction Hamiltonian due to the external field $A_{\mu}(\mf x)$ is obtained by integrating Eq.~\eqref{h-int-density} over the slice $t = \text{const}.$,
\begin{equation}\label{H-interaction}
    \hat{H}_{\text{ext}}(t) = -  q\int \dd^3 \bm x \,\hat{\bar{\psi}}(\mf{x})\slashed{A}(\mf x)\hat{\psi}(\mf{x})\, .
\end{equation}
Here, the quantum field $\hat{\psi}(\mf x)$ is given by Eq.~\eqref{psi-general}. The interaction Hamiltonian $\hat{H}_\text{ext}$ can be projected to the $\mathcal{H}_\text{s}$ subspace using the projector $\hat{P}_\text{s}$ from the previous section. It reads
\begin{align}\label{zeeman-non-reduced}
    &\hat{H}_I(t) \coloneqq \hat{P}_\text{s}\hat{H}_{\text{ext}}(t)\hat{P}_\text{s} \\
    &= -  q\sum_{m \in \{\uparrow, \downarrow\}} \sum_{ m' \in \{\uparrow, \downarrow\}}  \int \dd^3 \bm x\, \hat{\bar{\psi}}_m(\bm x)\slashed{A}(\mf x)\hat{{\psi}}_{m'}(\bm x)\ket{m}\bra{m'}\nonumber,
\end{align}
where the time dependence of the modes $\psi_{\uparrow}$ and $\psi_{\downarrow}$ cancels, as $E_\uparrow = E_\downarrow$. 
The states $\psi_\uparrow(\bm x)$ and $\psi_\downarrow(\bm x)$ are solutions to the time-independent Dirac equation:
\begin{align}
    \psi_\uparrow(\bm x) = \ii\begin{pmatrix}
    g(r)\frac{1}{2\sqrt{\pi}}\\
    0\\
    -\ii f(r)\frac{z}{2\sqrt{\pi}r}\\
    -\ii f(r)\frac{x+\ii y}{2\sqrt{\pi}r}
\end{pmatrix} \, ,\label{Psi-lzero-up}\\
    \psi_\downarrow(\bm x) = \ii\begin{pmatrix}
    0\\
    g(r)\frac{1}{2\sqrt{\pi}}\\
    -\ii f(r)\frac{x-\ii y}{2\sqrt{\pi}r}\\
    \ii f(r)\frac{z}{2\sqrt{\pi}r}\label{Psi-lzero-down}\\
\end{pmatrix} \, ,
\end{align}
where $\bm x = (x,y,z)$ and $r = |\bm x|$. Importantly, the mode functions $f(r)$ are significantly smaller in magnitude than the mode functions $g(r)$, with $f(r)/g(r) = \mathcal{O}(\alpha)$. For instance, when $n_0 = 1$, the radial functions $g$ and $f$ are
\begin{align}
    g(r) &= k_1a_0^{-3/2}e^{-Z r/a_0}(r/a_0)^{\beta-1}\label{g} \, ,\\*
    f(r) &= k_2a_0^{-3/2}e^{-Z r/a_0}(r/a_0)^{\beta-1}\label{f}\, ,
\end{align}
with $a_0$, $\beta$ $k_1$ and $k_2$ defined as
\begin{align}
    a_0 &= \frac{1}{m_e\alpha}\, , 
  &&&  \beta &= \sqrt{1 - Z^2\alpha^2}\label{a0beta} \, ,\\
    k_1 &= 2^\beta Z^{\beta + \tfrac{1}{2}} \sqrt{\frac{1+\beta}{\Gamma(1+2\beta)}} \, ,  &&&
    k_2 &= -k_1\sqrt{\frac{1-\beta}{1+\beta}},\label{k1k2}
\end{align}
 where $a_0$ is the Bohr radius. In this example one can see that the functions $f(r)$ are significantly smaller than $g(r)$ by noticing that  $k_2 = \mathcal{O}(\alpha)$, while $k_1$ is  $\mathcal{O}(1)$.

Substituting the solutions (\ref{Psi-lzero-up}), (\ref{Psi-lzero-down}) in the interaction Hamiltonian density (\ref{zeeman-non-reduced}), we obtain
\begin{align}\label{H-interaction-simple}
    \hat{H}_I(t) &=  q \int \dd^3\bm x\, \frac{f(r)g(r)}{2\pi r}\hat{\bm{\sigma}}\cdot(\bm{x}\times \bm{A}(t, \bm{x}))\, .
\end{align}
It is possible to rewrite $\hat{H}_I(t)$ as a smeared version of the Zeeman Hamiltonian~\eqref{eq:ZeemanBase} added to a boundary term by using the following vector identities 
\begin{align}
    \bm \nabla \psi {\times} \bm{v} &= \bm \nabla \times (\psi \bm{v} ) - \psi \bm \nabla {\times} \bm{v} \label{prop2}\, ,\\
    \bm u \cdot (\bm \nabla\times \bm v) & = (\bm \nabla \times \bm u) \cdot \bm v - \bm \nabla \cdot( \bm u \times \bm v) \label{prop3}\, .
\end{align}
Let us first define the useful function $\upphi(r)$ in terms of the radial functions $f(r)$ and $g(r)$ by the conditions
\begin{equation}\label{Phi-def}
    \bm{\nabla}\upphi(r) = \frac{f(r)g(r)}{r}\bm{x}\, ,\quad \lim_{r\to\infty} \upphi(r) = 0.
\end{equation}
Using $\bm \nabla \upphi(r) = \partial_r\upphi(r)\bm{x}/r$, the unique solution to Eq.~\eqref{Phi-def} can be written as\footnote{For instance, when $n_0 = 1$ we can write $\upphi(r)$ explicitly in terms of the incomplete gamma function $\Gamma(s,t)$: 
\begin{align}\label{Phi}
    \upphi(r) = \frac{2\sqrt{1-\beta^2}}{a_0^2 \Gamma(2\beta+1)} \, \Gamma(2\beta - 1, 2r/a_0) \, .
\end{align}},
\begin{equation}\label{eq:phir}
    \upphi(r) = -\int_r^\infty \dd r\, f(r)g(r) .
\end{equation}

We can then recast the interaction Hamiltonian as
\begin{equation}
    \hat{H}_I(t) = \frac{q}{2\pi} \int \dd^3\bm x\, \hat{\bm{\sigma}}\cdot(\nabla\upphi\times \bm{A}(t, \bm{x}))\, .
\end{equation}
Using (\ref{prop2}) we obtain a dependence on $\bm B = \nabla\times \bm A$:
\begin{align}
    \hat{\bm{\sigma}}\cdot(\bm{\nabla}\upphi \times \bm{A})&=  \hat{\bm{\sigma}}\cdot(\bm{\nabla}\times(\upphi\bm{A}) - \upphi\bm{\nabla}\times \bm{A})\nonumber\\
    &= - \upphi\,\hat{\bm{\sigma}}\cdot\bm{B} +  \hat{\bm{\sigma}}\cdot(\bm{\nabla} \times (\upphi\bm{A})).\label{eq:RuhiIsAStar}
\end{align}
where the term $\hat{\bm{\sigma}}\cdot(\bm{\nabla} \times (\upphi\bm{A}))$ can be recast as a boundary term using~\eqref{prop3}:
\begin{align}
    \hat{\bm{\sigma}}\cdot(\bm{\nabla} \times (\upphi\bm{A})) &=   - \bm{\nabla}\cdot(\hat{\bm{\sigma}} \times (\upphi\bm{A})) + (\bm{\nabla}\times\hat{\bm{\sigma}})\cdot \upphi\bm{A}\nonumber\\    
    &=  - \bm{\nabla}\cdot(\hat{\bm{\sigma}} \times (\upphi\bm{A})) \, ,
\end{align}
where we used $\nabla\times\hat{\bm \sigma} = 0$, given that $\hat{\bm \sigma}$ is independent of $\bm x$. The boundary term can be safely neglected due to the fact that $\upphi(r)$ decays exponentially as $r$ increases. 

Plugging the result of Eq.~\eqref{eq:RuhiIsAStar} in the Hamiltonian~\eqref{H-interaction-simple} and neglecting the boundary term, we obtain
\begin{equation}\label{H-interaction-zeeman}
    \hat{H}_I(t) = - \frac{q}{2\pi} \int d^3\bm x\, \upphi(r)\, \hat{\bm{\sigma}}\cdot\bm{B}(t, \bm{x})\, .
\end{equation}
The interaction Hamiltonian above can be seen as a generalization of the Zeeman effect, which takes into account that the magnetic field that couples to the spin is smeared by the function $\upphi(r)$.

The Zeeman interaction in its familiar form can be obtained by assuming that the magnetic field is approximately homogeneous within the localization of the atom in the same spirit as in the dipole approximation~\cite{ScullyBook}. Indeed, if $\bm B(t,\bm x) = \bm B(t)$ the radial integral of $\upphi(r)$ factors out. We compute this integral in Appendix~\ref{app:derivation}, resulting in
\begin{equation}\label{hflatzeemanexact}
\hat{H}_I(t) =  -\frac{q}{2m_e}\hat{\bm{\sigma}}\cdot \bm{B}(t)\left(1 -\frac{4}{3}\int_0^\infty \!\dd r \, r^2 f^2(r)\right).
\end{equation}
Using the fact that $f(r) = \mathcal{O}(\alpha)$, and defining the spin operator $\hat{\bm S} = \frac{\hbar}{2} \hat{\bm \sigma}$, we find
\begin{equation}\label{eq:HintB}
    \hat{H}_I(t) = - \frac{q}{m_e} \hat{\bm{S}}\cdot \bm{B}(t) + \mathcal{O}(\alpha^2) \, .
\end{equation}
To leading order in the fine-structure constant $\alpha$, this is the Zeeman Hamiltonian for the ground state splitting of a Hydrogen atom. The higher order corrections in $\alpha$ are defined by the specific shape of the mode functions $f(r)$. Equation~\eqref{H-interaction-zeeman} effectively gives the corrections to the electron $g$-factor due to the fact that the electron is localized in an atomic orbital. This effect has been first noted by Breit in the case of hydrogen-like atoms in~\cite{Breit1928}.

Notice that the leading order corrections in $\alpha$ to the Zeeman Hamiltonian presented above are of order $\mathcal{O}(\alpha^2)$, while it is well known that corrections from QED interactions to the electron $g$-factor are of first order in $\alpha$. To introduce these QED of corrections it would be enough to incorporate the renormalized interactions obtained from higher loop QED considerations~\cite{gfactorBeier2000,gfactorIndelicato2004,gfactorTwoLoop2013,gfactor2020,gfactorExp2023}.

Finally, notice that the reduction of the QED interaction Hamiltonian to Eq.~\eqref{H-interaction-zeeman} presented above is also valid when the electron field is under the influence of any spherically symmetric electric potential\footnote{This would allow one to consider corrections to the Coulomb potential due to the finite size of the proton, or due to electrons in inner shells.} $A_0(r)$. This can be seen by noticing that localized mode solutions with quantum numbers $j=1/2$ and $p=+1$ take the form of Eqs.~\eqref{Psi-lzero-up} and~\eqref{Psi-lzero-down} with different radial functions $f(r)$ and $g(r)$ determined by $A_0(r)$.


\section{A spin as a local probe of the quantum electromagnetic field}\label{sec:quantumProbe}

The past few decades have seen the development of  many quantum information protocols (i.e., practically useful tasks whose performance takes advantage of the quantum nature of matter) that make explicit use of the quantum degrees of freedom of relativistic quantum fields. Examples of such protocols are entanglement harvesting~\cite{Valentini1991,Reznik1,reznik2,Salton:2014jaa,Pozas-Kerstjens:2015,HarvestingQueNemLouko,Pozas2016,HarvestingSuperposed,Henderson2019,bandlimitedHarv2020,ampEntBH2020,mutualInfoBH,threeHarvesting2022,twist2022,quantClass}, quantum collect calling~\cite{Jonsson2,collectCalling,PRLHyugens2015,Simidzija_2020}, and quantum energy teleportation~\cite{teleportation,teleportation2014,nichoTeleport,teleportExperiment}. These have a feature in common: they all require local probes that couple to a quantum field in a finite region of spacetime. These probes are often\footnote{Note that the protocols have also been studied for more involved probes~\cite{generalPD}, such as harmonic-oscillator detectors~\cite{RaineHO1991,HOdetectorsBeiLok1994,Hu_Jorka_HO_2012,Bruschi_HO2013,Brown_HO2013}, non-relativistic hydrogen-like atoms dipolarly coupled to the electric field~\cite{eduardoOld2013,Pozas2016}, multipole extensions of this coupling~\cite{richard}, quantum systems coupled to the linearized gravitational field~\cite{boris}, and nuclei coupled to the neutrino field~\cite{neutrinos,carol}.} modelled as two-level Unruh-DeWitt (UDW) detectors~\cite{Unruh1976,DeWitt} that couple to a scalar quantum field $\hat{\phi}(\mf x)$. In this simplified model, the detector is a qubit with internal energy gap $\Omega$ that undergoes a trajectory in spacetime. If we consider such a detector in an inertial trajectory given by $(t,\bm 0)$, the interaction with the field is given by the Hamiltonian density
\begin{equation}\label{eq:hiUDW}
    \hat{h}_I(\mf x) = \lambda \hat{\sigma}_x(t) \chi(t) f(\bm x) \hat{\phi}(\mf x),
\end{equation}
where $\lambda$ is a dimensionless coupling constant, $\chi(t)$ is a switching function that determines the time profile of the interaction, $f(\bm x)$ is a smearing function that determines the shape of the detector (and the spatial profile of the interaction), and $\hat{\sigma}_x(t) = e^{\ii \Omega t}\hat{\sigma}_+ + e^{- \ii \Omega t}\hat{\sigma}_-$ is the time-evolved monopole operator of the detector. Despite its simplicity, this model captures the fundamental aspects of the interaction of localized probes with quantum fields~\cite{eduardoOld2013,antiparticles}.

In this section we will show that the model of Section~\ref{sec:coupleExtMag} for the coupling of a spin with an external magnetic field can be used to locally probe the quantum magnetic field, in analogy to the UDW detector model. In this context, the electron in the atom is used as the probe, described as a localized mode of the fermionic field $\hat{\psi}(\mf x)$. This is akin to the description first conceived by Unruh~\cite{Unruh1976}, and later explored in~\cite{QFTPD,FullHarvesting}, where the localized probe is itself modelled by a quantum field.

We will analyze two situations: first the case where the electron energy is degenerate in spin, and then a case where an external constant field splits the energy of the different spin configurations. The latter analysis will allow for a better comparison with the usual UDW models.

\subsection{A degenerate spherically symmetric spin system interacting with an external electromagnetic field}

We consider the effective two-level description of Section~\ref{sec:coupleExtMag}, so that the electron is coupled to an external quantum electromagnetic field according to the interaction Hamiltonian of Eq.~\eqref{eq:HintB}. We will go through the calculations in detail in this section for pedagogical purposes, highlighting the overall steps and identities relevant to the model. We reserve Appendix~\ref{app:the full dynamics} to the more general computations.

The first step for a relativistic description of a spin as a probe of the electromagnetic field is to describe the external field $A_\mu(\mf x)$ in the framework of QFT. Being consistent with our gauge choice used to solve for the modes of the fermionic field $\hat{\psi}(\mf x)$ in Section~\ref{sec:QED-Atom}, we write the quantized electromagnetic potential $\hat{A}_\mu(\mf x)$ in the Coulomb gauge: 
\begin{equation}
    \hat{\bm A}(\mf x) =\sum_{s=1}^2\int \frac{\dd^3\bm k}{\sqrt{2|\bm k|}}\left(\frac{e^{\ii \mf k \cdot \mf x}}{(2\pi)^{\frac{3}{2}}}\hat{a}_{\bm k,s} + \frac{e^{-\ii \mf k \cdot \mf x}}{(2\pi)^{\frac{3}{2}}}\hat{a}_{\bm k,s}^\dagger\right) \bm{\mathcal{E}}_s(\bm k),
\end{equation}
where $\mf k \cdot \mf x = \eta_{\mu\nu} k^\mu x^\nu$, 
 with $\eta_{\mu\nu} = \text{diag}(-1,1,1,1)$, \mbox{$\mf k = (|\bm k|,\bm k)$}, \mbox{$\mf x = (t,\bm x)$}. The vectors $\bm{\mathcal{E}}_s(\bm k)$ are orthonormal transverse polarization vectors and $\hat{a}_{\bm k,s}^\dagger$ and $\hat{a}_{\bm k,s}$ are the creation and annihilation operators for the modes with momentum $\bm k$ and polarization $s$, satisfying the canonical commutation relations
\begin{equation}
    [\hat{a}_{\bm k,s},\hat{a}_{\bm k',s'}^\dagger] = \delta_{ss'} \delta^{(3)}(\bm k - \bm k').
\end{equation}
We can explicitly parametrize the set of $\bm k$'s, as well as the polarization vectors in terms of the parameters $|\bm k|\in (0,\infty)$, $\theta\in(0,\pi)$ and $\phi\in(0,2\pi)$:
\begin{gather}
    \bm k = |\bm k|(\sin\theta\cos\phi\bm e_x + \sin\theta \bm e_y + \cos\theta \bm e_z),\\
    \bm{\mathcal{E}}_1(\bm k) = \cos\theta\cos\phi \bm e_x + \cos\theta \sin\phi \bm e_y - \sin\theta \bm e_z,\label{eq:e1}\\
    \bm{\mathcal{E}}_2(\bm k) = -\sin\phi \bm e_x + \cos\phi \bm e_y.\label{eq:e2}
\end{gather}
We obtain the quantum description of the magnetic field in the frame $(t,\bm x)$ by taking the curl of $\hat{\bm A}(\mf x)$:
\begin{equation}
    \hat{\bm B}(\mf x) = \nabla\times\hat{\bm A}(\mf x),
\end{equation}
which can be explicitly written as
\begin{equation}\label{eq:Bexpansion}
    \hat{\bm B}(\mf x) = \ii\sum_{s=1}^2\int \!\!\frac{\dd^3\bm k}{\sqrt{2|\bm k|}}\!\left(\!\frac{e^{\ii \mf k \cdot \mf x}}{(2\pi)^{\frac{3}{2}}}\hat{a}_{\bm k,s} - \frac{e^{-\ii \mf k \cdot \mf x}}{(2\pi)^{\frac{3}{2}}}\hat{a}_{\bm k,s}^\dagger\!\right)\!|\bm k| \bm\epsilon_s(\bm k),
\end{equation}
where $\bm{\epsilon}_s(\bm k)$ are the orthonormal vectors
\begin{align}
    \bm \epsilon_1(\bm k) &\coloneqq \bm k \times \bm{\mathcal{E}}_1(\bm k)/|\bm k| = \bm{\mathcal{E}}_2(\bm k), \\ 
    \bm \epsilon_2(\bm k) &\coloneqq \bm k \times \bm{\mathcal{E}}_2(\bm k)/|\bm k| =  - \bm{\mathcal{E}}_1(\bm k).
\end{align}

The localized interaction between the electron spin (the detector) and the magnetic field is prescribed by the Hamiltonian of Eq.~\eqref{eq:HintB} using the quantum field $\hat{\bm B}(\mf x)$ and adding a switching function $\chi(t)$ that localizes the interaction in time\footnote{Notice that the addition of a switching function is merely a generalization of the interaction, so that the limit of constant interactions can be recovered by taking $\chi\to 1$, as well as finite time interactions when $\chi(t)$ is a window function.},
\begin{equation}
    \hat{H}_I(t) = - \frac{q}{2 \pi} \chi(t) \int \dd^3 \bm x \,\upphi(|\bm x|)\,\hat{\bm \sigma}\cdot \hat{\bm B}(\mf x).
\end{equation}
The interaction Hamiltonian density is then
\begin{equation}\label{eq:hIspinEMdetector}
    \hat{h}_I(\mf x) = - \frac{q}{2 \pi} \chi(t) \upphi(|\bm x|)\,\hat{\bm \sigma}\cdot \hat{\bm B}(\mf x).
\end{equation}
Notice that the spatial localization of the interaction is given by the function ${\upphi(|\bm x|)}$, defined in~\eqref{Phi} by the localization of the spinor modes. For convenience, we define the spacetime smearing function
\begin{equation}\label{eq:spacetimeSmearingAtom}
    \Lambda(\mf x) \coloneqq \chi(t)\upphi(|\bm x|),
\end{equation}
which determines the shape of the interaction region in spacetime.

Time evolution of the spin-field system due to the interaction is implemented by the unitary time evolution operator
\begin{equation}
    \hat{U}_I = \mathcal{T}\exp\left(\int \dd V \hat{h}_I(\mf x) \right),
\end{equation}
where $\mathcal{T}\exp$ denotes the time-ordered exponential. To compute $\hat{U}_I$ and the effect of the interaction with the field on the qubit, we can work perturbatively in the coupling constant $q$, writing the time evolution operator as the Dyson series 
\begin{equation}\label{eq:DysonBase}
    \hat{U}_I = \openone + \hat{U}_I^{(1)} + \hat{U}_I^{(2)} + \mathcal{O}(q^3),
\end{equation}
where
\begin{align}
    \hat{U}_I^{(1)} &= - \ii \int \dd V \hat{h}_I(\mf x),\label{eq:Dyson1}\\
    \hat{U}^{(2)}_I &= - \int \dd V \dd V' \hat{h}_I(\mf x) \hat{h}_I(\mf x') \theta(t-t'),\label{eq:Dyson2}
\end{align}
and $\theta(u)$ denotes the Heaviside theta function.

We consider that the field and detector state start uncorrelated, so that the initial state of the system is $\hat{\rho}_0 = \hat{\rho}_{\tc{d},0}\otimes \hat{\rho}_{B}$, where $\hat{\rho}_B = \ket{0}\!\bra{0}$ is the vacuum state of the electromagnetic field, and
\begin{equation}\label{eq:rhoBloch}
    \hat{\rho}_{\tc{d},0} = \frac{1}{2}\left(\openone + \bm a \cdot \bm \sigma\right)
\end{equation}
is a general qubit state, with $|\bm a|\leq 1$. Applying the time evolution operator to the initial state $\hat{\rho}_0$ allows one to obtain the final state of the detector by tracing over the degrees of freedom of the electromagnetic field,
\begin{align}\label{eq:rho2Dyson}
    \hat{\rho}_\tc{d} =& \tr_B\left(\hat{U}_I \hat{\rho}_{0}\hat{U}_I^\dagger\right),
\end{align}
where
\begin{align}
    \hat{U}_I \hat{\rho}_{0}\hat{U}_I^\dagger = & \hat{\rho}_{\tc{d},0} + \hat{U}_I^{(1)} \hat{\rho}_{0} + \hat{\rho}_{0}\hat{U}_I^{(1)\dagger}\\& + \hat{U}_I^{(2)} \hat{\rho}_{0} +  \hat{U}_I^{(1)} \hat{\rho}_{0}\hat{U}_I^{(1)\dagger} +   \hat{\rho}_{0}\hat{U}_I^{(2)\dagger}\nonumber + \mathcal{O}(q^3).
\end{align}
Notice that the electromagnetic vacuum $\ket{0}$ is a Gaussian state with zero mean (commonly referred to as a quasifree state~\cite{Haag}), implying that ${\tr_B(\hat{U}_I^{(1)} \hat{\rho}_{0})}$ and ${\tr_B(\hat{\rho}_{0}\hat{U}_I^{(1)\dagger})}$ identically vanish due to the zero expectation value of the field in the vacuum.

We proceed to compute the remaining terms in Eq.~\eqref{eq:rho2Dyson}:
\begin{align}
    \tr_B\left( \hat{U}_I^{(1)} \hat{\rho}_{0}\hat{U}_I^{(1)\dagger}\right) &= \left(\frac{q}{2\pi}\right)^{\!2}\int  \dd V \dd V'\Lambda(\mf x) \Lambda(\mf x')\label{eq:U1rhoU1}\\
    & \:\:\:\:\:\:\:\:\:\:\:\:\:\:\:\:\:\:\:\:\:\times\sigma_i\hat{\rho}_{\tc{d},0} \hat{\sigma}_j \langle \hat{B}^j(\mf x')\hat{B}^i(\mf x) \rangle,\nonumber\\
    \tr_B\left( \hat{U}_I^{(2)} \hat{\rho}_{0}\right)\label{eq:U2rho}&= \!-\left(\frac{q}{2\pi}\right)^{\!2}\!\!\!\int\dd V \dd V' \Lambda(\mf x) \Lambda(\mf x')\\
    & \:\:\:\:\:\:\times\hat{\sigma}_i \hat{\sigma}_j \hat{\rho}_{\tc{d},0} \langle \hat{B}^i(\mf x) \hat{B}^j(\mf x')\rangle \theta(t-t'),\nonumber\\
    \tr_B\left(\hat{\rho}_{0}\hat{U}_I^{(2)\dagger}\right) 
    &= \!-\left(\frac{q}{2\pi}\right)^{\!2}\!\!\!\int\dd V \dd V' \Lambda(\mf x) \Lambda(\mf x')  
 \label{eq:rhoU2}\\
    &\:\:\:\:\:\:\times\hat{\rho}_{\tc{d},0}\hat{\sigma}_i \hat{\sigma}_j \langle \hat{B}^i(\mf x) \hat{B}^j(\mf x')\rangle \theta(t'-t)\nonumber,
\end{align}
where the expected values above are taken in the vacuum state $\ket{0}$:

\begin{equation}\label{eq:Bpropag}
    \langle \hat{B}^i(\mf x) \hat{B}^j(\mf x')\rangle = \frac{1}{(2\pi)^3}\sum_{s = 1}^2\int \frac{\dd^3 \bm k}{2|\bm k|}|\bm k|^2 e^{\ii\mf k\cdot (\mf x - \mf x')}\bm \epsilon_s^i(\bm k)\bm \epsilon_s^j(\bm k).
\end{equation}
By noticing that the propagator in Eq.~\eqref{eq:Bpropag} is symmetric in $i$ and $j$, we see that only the symmetric part of the tensors $\hat{\sigma}_i\hat{\sigma}_j \hat{\rho}_{\tc{d},0}$, $\hat{\sigma}_i \hat{\rho}_{\tc{d},0}\hat{\sigma}_j$, and $\hat{\rho}_{\tc{d},0}\hat{\sigma}_i\hat{\sigma}_j$ is relevant to the calculation. We can write these terms by using Eq.~\eqref{eq:rhoBloch} and $\hat{\sigma}_i \hat{\sigma}_j = \delta_{ij} \openone + \ii \epsilon_{ijk}\hat{\sigma}^k$, leading to
\begin{align}
    \hat{\sigma}_{(i}\hat{\sigma}_{j)} \hat{\rho}_{\tc{d},0} &= \tfrac{1}{2}(\openone +a^k \hat{\sigma}_k)\delta_{ij} = \hat{\rho}_{\tc{d},0}\delta_{ij},\label{eq:sisjrho}\\
    \hat{\sigma}_{(i} \hat{\rho}_{\tc{d},0}\hat{\sigma}_{j)} &= \tfrac{1}{2}( (\openone - a^k \hat{\sigma}_k)\delta_{ij} + 2 a_{(i}\hat{\sigma}_{j)}),\label{eq:sirhosj}\\
    \hat{\rho}_{\tc{d},0}\hat{\sigma}_{(i}\hat{\sigma}_{j)} &= \tfrac{1}{2}(\openone + a^k \hat{\sigma}_k)\delta_{ij} = \hat{\rho}_{\tc{d},0}\delta_{ij},\label{eq:rhosisj}
\end{align}
where we use the standard notation for symmetrized indices, $A_{(ij)} = \tfrac{1}{2} (A_{ij} + A_{ji})$. 

With the results of Eqs.~\eqref{eq:sisjrho} and \eqref{eq:rhosisj}, the terms $\langle \hat{U}_I^{(2)} \hat{\rho}_{0}\rangle_B$ and $\langle  \hat{\rho}_{0}\hat{U}_I^{(2)\dagger}\rangle_B$ combine and we can write
\begin{align}
    \langle \hat{U}_I^{(2)} &\hat{\rho}_{0}\rangle_B + \langle  \hat{\rho}_{0}\hat{U}_I^{(2)\dagger}\rangle_B\\
    &=  -\frac{q^2}{(2\pi)^5}\hat{\rho}_{\tc{d},0}\int \dd^3\bm k |\bm k|^3 |\tilde{\chi}(|\bm k|)|^2|\tilde{\upphi}(\bm k)|^2\nonumber,
\end{align}
where we defined the Fourier transforms
\begin{align}
    \tilde{\upphi}(\bm k) &= \int \dd^3 \bm x\, \upphi(\bm x) e^{\ii \bm k \cdot \bm x},\\
    \tilde{\chi}(\omega) &= \int \dd t \chi(t) e^{\ii \omega t}.
\end{align}
and used the completeness relation $\delta_{ij}\epsilon_s^i(\bm k)\epsilon_s^j(\bm k) = 1$ for $s = 1,2$. We can further expand the equation above by performing the integration over the $\bm k$ angles, $\vartheta$ and $\varphi$. Using spherical symmetry of $\upphi(|\bm x|)$, we obtain
\begin{align}
    \langle \hat{U}_I^{(2)} \hat{\rho}_{0}\rangle_B + \langle  \hat{\rho}_{0}\hat{U}_I^{(2)\dagger}\rangle_B=  -\left(\frac{q}{2\pi}\right)^{\!2} 2\pi (\openone + a^k \hat{\sigma}_k)\mathcal{L},\label{eq:02and20}
\end{align}
where
\begin{equation}
    \mathcal{L} = \frac{1}{(2\pi)^3}\int \dd |\bm k| |\bm k|^3 |\tilde{\chi}(|\bm k|)|^2|\tilde{\upphi}(|\bm k|)|^2,
\end{equation}
and the Fourier transform of $\upphi(|\bm x|)$ only depends on the norm $|\bm k|$ due to spherical symmetry.

The term $\langle \hat{U}_I^{(1)} \hat{\rho}_{0}\hat{U}_I^{(1)\dagger}\rangle_B$ involves contractions of the polarization tensors with $a_{(i}\hat{\sigma}_{j)}$, which give different factors for the angular integrals. Using Eqs.~\eqref{eq:Bpropag}, and~\eqref{eq:sirhosj} in Eq.~\eqref{eq:U1rhoU1}, we find
\begin{align}
    \langle \hat{U}_I^{(1)} \hat{\rho}_{0}&\hat{U}_I^{(1)\dagger}\rangle_B\label{eq:11}\\
    & = \left(\frac{q}{2\pi}\right)^{\!2} \left(2\pi (\openone - a^k \hat{\sigma}_k) + \frac{4\pi}{3} a^k \hat{\sigma}_k\right)\mathcal{L}.\nonumber
\end{align}
Adding Eqs.~\eqref{eq:02and20} and~\eqref{eq:11}, the identity terms cancel and we find that the final state of the detector can be written as
\begin{equation}\label{eq:rhoDgapless}
    \hat{\rho}_\tc{d} = \hat{\rho}_{\tc{d},0} - \frac{8\pi}{3} \left(\frac{q}{2\pi}\right)^{\!2} \mathcal{L}\, a^k \hat{\sigma}_k + \mathcal{O}(q^4)
\end{equation}
That is, if the qubit starts in a state with Bloch vector $\bm a$, after the interaction with the field, the leading order qubit's state is determined by the vector $\bm a + \delta \bm a$, where
\begin{equation}
    \delta \bm a = - \frac{16\pi}{3}\left(\frac{q}{2\pi}\right)^{\!2} \mathcal{L} \,\bm a.
\end{equation}
Thus, the qubit's final state in the Bloch sphere tends towards its center, as the variation of its Bloch vector is negatively proportional to itself. This is to be expected, as we are considering a gapless system, which only experiences noise through its interaction with the magnetic field. This will not be the case in the next Subsection when we consider an energy splitting between the spin states.

\subsection{A spin in an external magnetic field interacting with the electromagnetic vacuum}\label{sub:gapped}

While the interaction described above allows a spin to interact and probe an external quantum electromagnetic field, we considered a degenerate spin state, which makes them unable to access some relevant information about the field. For instance, at leading order a gapless detector cannot distinguish the vacuum from any Fock wavepacket (or any state with zero one-point function) in the adiabatic limit of long interaction times. However, one can adapt the spin model by considering an external  (classical) constant magnetic field that creates an energy gap for the detector. For instance, already in the case of the hydrogen atom, the effective magnetic field generated by the nucleus would give rise to a finite energy gap for the states of spin up and spin down, corresponding to the hyperfine structure. 

In order to produce an energy gap for the spin system, we consider a constant external electromagnetic field aligned with the $z$ axis with strength $|\bm B_0|$. The total magnetic field can the be expressed as
\begin{equation}
    \hat{\bm B}_{\text{total}}(\mf x) = \bm B_0 + \hat{\bm B}(\mf x),
\end{equation}
where $\hat{\bm B}(\mf x)$ incorporates the quantum degrees of freedom of the electromagnetic field, described by the expansion of Eq.~\eqref{eq:Bexpansion}. The total Hamiltonian density for the spin system then becomes
\begin{align}
     \hat{h}_I(\mf x) &= -\frac{q}{2\pi} \upphi(|\bm x|) \,\hat{\bm \sigma}\cdot \hat{\bm B}_{\text{total}}(\mf x) \\
     &= - \frac{q}{2\pi} \upphi(|\bm x|) |\bm B_0| \hat{\sigma}_z - \frac{q}{2\pi} \upphi(|\bm x|) \hat{\bm \sigma}\cdot \hat{\bm B}(\mf x).
\end{align}
For convenience, we define
\begin{equation}
    \Omega \coloneqq - \frac{q}{\pi}|\bm B_0|\int \dd^3 \bm x\, \upphi(|\bm x|), \quad\quad \hat{H}_\tc{d} \coloneqq \frac{\Omega}{2} \hat{\sigma}_z
\end{equation}
so that $|\Omega|$ corresponds to the energy gap between the two spin states $\ket{\uparrow}$ and $\ket{\downarrow}$, generated by the classical external magnetic field. Namely, the Hamiltonian $\hat{H}_\tc{d}$ defines the excited state $\ket{\downarrow}$, and the ground state $\ket{\uparrow}$ \footnote{due to the negative charge of the electron we have $\Omega<0$.}. 

We can then move to the interaction picture by incorporating the time evolution generated by $\hat{H}_\tc{d}$ to the qubit observables. We then have
\begin{align}
    \hat{\sigma}_x(t) &= \cos(\Omega t) \hat{\sigma}_x - \sin(\Omega t) \hat{\sigma}_y \label{eq:sigmaxt}\\
    \hat{\sigma}_y(t) &= \sin(\Omega t) \hat{\sigma}_x + \cos(\Omega t) \hat{\sigma}_y\label{eq:sigmayt}\\
    \hat{\sigma}_z(t) &= \hat{\sigma}_z.
\end{align}
We also implement a switching function $\chi(t)$ to the interaction with the quantum external electromagnetic field so that the interaction Hamiltonian density in the interaction picture can be written as
\begin{equation}\label{eq:sphsymmUDW}
    \hat{h}_I(\mf x) = - \frac{q}{2\pi} \Lambda(\mf x)\,\hat{\bm \sigma}(t)\cdot \hat{\bm B}(\mf x),
\end{equation}
with $\Lambda(\mf x)$ as in Eq.~\eqref{eq:spacetimeSmearingAtom}. 

The time dependence introduced in Eqs.~\eqref{eq:sigmaxt} and~\eqref{eq:sigmayt} make the dynamics more involved. Under the assumption of an uncorrelated initial state for the detector and electromagnetic field, one can still obtain the Dyson expansion expressions in Eqs~\eqref{eq:U1rhoU1}, \eqref{eq:U2rho}, and \eqref{eq:rhoU2}, with the replacement $\hat{\sigma}_i \mapsto \hat{\sigma}_i(t)$. We explicitly compute the leading order final state of the detector in Appendix~\ref{app:the full dynamics}. It can be written as
\begin{align}
    \hat{\rho}_\tc{d} &\,= \hat{\rho}_{\tc{d},0}\label{eq:rhoGapSpin}\\
    &- \frac{4\pi}{3} \!\left(\frac{q}{2\pi}\right)^{\!2}\!\!\big(a^x (\mathcal{L}(0) + \Re\mathcal{M}(\Omega)) + a^y\Im\mathcal{M}(\Omega)\big)\hat{\sigma}_x\nonumber\\
    &- \frac{4\pi}{3} \!\left(\frac{q}{2\pi}\right)^{\!2}\!\!\big(a^y(\mathcal{L}(0) + \Re\mathcal{M}(\Omega)) - a^x\Im\mathcal{M}(\Omega)\big)\hat{\sigma}_y\nonumber\\ 
    &- \frac{4\pi}{3} \!\left(\frac{q}{2\pi}\right)^{\!2}\!\!\big(a^z (\mathcal{L}(\Omega) + \mathcal{L}(-\Omega)) - {\mathcal{L}}(\Omega)+ {\mathcal{L}}(-\Omega)\big)\hat{\sigma}_z\nonumber,
\end{align}
where
\begin{align}
    \mathcal{L}(\Omega) &= \frac{1}{(2\pi)^3}\int \dd |\bm k| |\bm k|^3 |\tilde{\upphi}(|\bm k|)|^2 |\tilde{\chi}(|\bm k| + \Omega)|^2,\label{eq:Lomega}\\
    \mathcal{M}(\Omega) & = \frac{1}{(2\pi)^3}\int \dd |\bm k| |\bm k|^3 |\tilde{\upphi}(|\bm k|)|^2 Q(|\bm k|,\Omega),\label{eq:Momega}
\end{align}
with
\begin{equation}\label{eq:Qkomega}
    Q(|\bm k|,\Omega) = \int \dd t \dd t' \chi(t)\chi(t') e^{\ii \Omega |t-t'|}e^{- \ii |\bm k|(t-t')}.
\end{equation}

In the final state of Eq.~\eqref{eq:rhoGapSpin}, the $\mathcal{L}(\pm\Omega)$ terms are associated with spin transitions between ground and excited states (spin flips), and the $\mathcal{M}(\Omega)$ terms determine a rotation of the Bloch vector in the $xy$ plane, which can be seen by noticing that
\begin{align}
    &\big(a^x  \Re\mathcal{M}(\Omega) + a^y\Im\mathcal{M}(\Omega)\big)\hat{\sigma}_x \nonumber\\*
    &+ \big(a^y\Re\mathcal{M}(\Omega) - a^x\Im\mathcal{M}(\Omega)\big)\hat{\sigma}_y \\*
    &= |\mathcal{M}(\Omega)|e^{-\ii \varphi (\Omega) \hat{\sigma}_z/2}(a^x \hat{\sigma}_x + a^y \hat{\sigma}_y)e^{\ii \varphi(\Omega) \hat{\sigma}_z/2}\nonumber
\end{align}
where $\varphi(\Omega) = \text{arg}(\mathcal{M}(\Omega))$.
The $\mathcal{L}(0)$ terms in the $x$ and $y$ components of Eq.~\eqref{eq:rhoGapSpin} are reminiscent from the local noise experienced in the gapless interaction~\eqref{eq:rhoDgapless}. To recover the degenerate result of Eq.~\eqref{eq:rhoDgapless} one can set $\Omega = 0$ in Eq.~\eqref{eq:rhoGapSpin} so that $Q(|\bm k|,0) = |\tilde{\chi}(|\bm k|)|^2$ implies $\mathcal{M}(0) = \mathcal{L}(0)\in\mathbb{R}$.

Importantly, the results of Eq.~\eqref{eq:rhoGapSpin} are time translation invariant. Indeed, shifting the interaction time by $t_0$ through the operation $\chi(t)\mapsto \chi(t-t_0)$ does not affect $|\tilde{\chi}(|\bm k|)|^2$ or $Q(|\bm k|,|\Omega|)$ so that $\mathcal{L}(\Omega)$ and $\mathcal{M}(\Omega)$ are invariant. This means that regardless of the initial state of the qubit (determined by the Bloch vector $\bm a$) and of the choice of initial time for the interaction, the same parameters $\mathcal{L}(\pm\Omega)$ and $\mathcal{M}(\Omega)$ determine the leading order time evolved state $\hat{\rho}_\tc{d}$ by Eq.~\eqref{eq:rhoGapSpin}.

We can now focus on a specific example of this interaction, when the initial state of the spin system starts in an eigenstate of the free Hamiltonian $\hat{H}_\tc{d}$. In this case, the Bloch vector of the spin is $\bm a = \pm \bm e_z$, corresponding to excited state (spin down, $a_z = -1$) and ground state (spin up, $a_z = 1$). The probability to flip the spin after a finite interaction with the electromagnetic field can be computed from Eq.~\eqref{eq:rhoGapSpin} and reads
\begin{equation}
    P_\text{flip}(\Omega) = \frac{8\pi}{3}\left(\frac{q}{2\pi}\right)^2 \mathcal{L}(\mp\Omega),
\end{equation}
where the sign of $\mp \Omega$ determines whether the state started in spin up, or in spin down ($a_z=\pm1\Rightarrow \mp\Omega$). Notice that even if the spin starts its interaction in the ground state, the external time dependence added by the switching function $\chi(t)$ may add energy into the system, allowing the qubit to become excited after its interaction. 

\begin{figure}[h!]
    \centering
    \includegraphics[width=8.6cm]{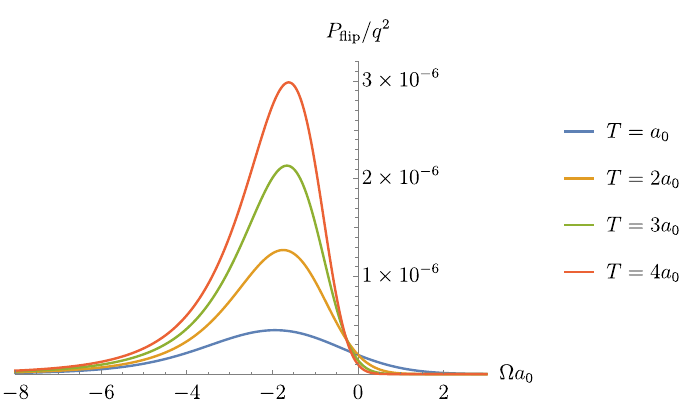}
    \caption{The excitation/deexcitation probability of an electron in the $1s$ orbital of a hydrogen atom after interacting with the vacuum of the quantum electromagnetic field as a function of $\Omega a_0$ with interaction times controlled by the time parameter $T$ in Eq.~\eqref{eq:chiGauss}.}
    \label{fig:pOmega}
\end{figure}

In Fig.~\ref{fig:pOmega} we display the transition probability of a hydrogen atom in the $1s$ orbital as a function of $\Omega$ with the choice of switching function
\begin{equation}\label{eq:chiGauss}
    \chi(t) = e^{-\frac{\pi t^2}{2T^2}},
\end{equation}
where $T$ is a parameter with units of time that controls the duration of the interaction. Despite perhaps being unusual for those used to particle detector calculations, we include the geometric factors in $\chi(t)$ as in Eq.~\eqref{eq:chiGauss} to match the conventions outlined in Appendix A of~\cite{JormaAndI}. The plot in Fig.~\ref{fig:pOmega} shows the standard behaviour for a quantum system coupled to a quantum field for a finite time, peaking when $\Omega$ is negative of the order of the inverse of the system's size (determined by $a_0$ in a hydrogen atom).

One can also obtain the transition rate of the qubit by considering the limit when the interaction is  switched on indefinitely (i.e., the adiabatic limit $T \to \infty$). Because this is the adiabatic limit, the transition rate can be written as the limit
\begin{equation}
    \dot{P}_\text{flip}(\Omega) = \lim_{T\to\infty} \frac{P_{\text{flip}}(\Omega)}{T}
\end{equation}
The transition rate can then be obtained by noticing that 
\begin{equation}
    \lim_{T\to\infty} \frac{|\tilde{\chi}(|\bm k|)|^2}{T} = 2\pi \delta(|\bm k| + \Omega).
\end{equation}
Setting $|\tilde{\chi}(|\bm k|)|^2 \mapsto 2\pi \delta(|\bm k| + \Omega)$ in Eq.~\eqref{eq:Lomega}, which yields
\begin{gather}
    \dot{P}(\Omega) = -\frac{8q^2 (1-\beta^2) \left(\sin(\omega) - \beta a_0 \Omega \cos(\omega)\right)^2}{3 \pi \beta^2 a_0^4\Omega^3 (2\beta - 1)\left(1+\frac{a_0^2 \Omega^2}{4}\right)^{2\beta}} \theta(- \Omega),
\end{gather} 
with $\omega = 2\beta \arctan(\tfrac{a_0\Omega}{2})$.








\subsection{Comparing a spin coupling to the magnetic field to a two-level UDW detector}

In this section we will compare the commonly employed UDW model (defined by the interaction of Eq.~\eqref{eq:hiUDW}) with the model of a spin interacting with the magnetic field discussed above. By comparing the interaction Hamiltonians~\eqref{eq:hiUDW} and~\eqref{eq:sphsymmUDW}, we see that the role of the coupling constant is played by $\lambda$ in the UDW model and by $q/2\pi$ in the spin interaction. Analogously, the spatial profile in the interaction is defined by the smearing function $f(\bm x)$ in the UDW model, while it is defined by $\upphi(|\bm x|)$ for the spin interacting with the magnetic field. Importantly, $f(\bm x)$ has units of volume density\footnote{Notice that we are working in $3+1$ dimensions, so that $f(\bm x)$ has units of energy cubed.}, while $\upphi(|\bm x|)$ has units of energy squared. This difference is due to the different units of a scalar quantum field, $\hat{\phi}(\mf x)$ (with units of energy) and the magnetic field, $\hat{\bm B}(\mf x)$ (with units of energy squared). The role of the switching function and of the energy gap are unchanged between the models.

Perhaps a more fair comparison to a scalar model for the coupling of spin with a magnetic field would be the so-called \textit{derivative coupling} UDW detector, which couples to $\partial_t\hat{\phi}(\mf x)$ instead of $\hat{\phi}(\mf x)$. If we were to consider this alternative model, the interaction Hamiltonian density would read
\begin{equation}\label{eq:hiUDW}
    \hat{h}_I(\mf x) = \lambda \hat{\sigma}_x(t) \chi(t) f(\bm x) \partial_t\hat{\phi}(\mf x).
\end{equation}
In this case, the units of the smearing function $f(\bm x)$ would be the same as those of $\upphi(|\bm x|)$, as the units of $\partial_t\hat{\phi}(\mf x)$ match those of $\hat{\bm B}(\mf x)$. 

To compare the models, we compute the evolution of a general qubit state that interacts with a scalar quantum field according to the UDW model. We consider the interaction Hamiltonian of Eq.~\eqref{eq:hiUDW} with a spherically symmetric smearing function $f(|\bm x|)$ and the initial qubit state of Eq.~\eqref{eq:rhoBloch}. To leading order in $\lambda$, the final state of a UDW detector after its interaction with the Minkowski vacuum of the massless field $\hat{\phi}(\mf x)$ can be written as
\begin{align}
    &\hat{\rho}_\tc{d} = \hat{\rho}_{\tc{d},0}\label{eq:rhoDUDW}\\
    &- \!\pi \lambda^2\big(a^x \Re(\overline{\mathcal{M}}(\Omega) \!-\! \overline{\mathcal{K}}(\Omega)) + a^y\Im(\overline{\mathcal{M}}(\Omega) \!+\! \overline{\mathcal{K}}(\Omega))\big)\hat{\sigma}_x\nonumber\\
    & - \!\pi \lambda^2 \big(a^y\Re(\overline{\mathcal{M}}(\Omega) \!+\! \overline{\mathcal{K}}(\Omega)) - a^x \Im(\overline{\mathcal{M}}(\Omega) \!-\! \overline{\mathcal{K}}(\Omega))\big)\hat{\sigma}_y\nonumber\\
    &-\!\pi \lambda^2 \big(a^z (\overline{\mathcal{L}}(\Omega) + \overline{\mathcal{L}}(-\Omega)) - \overline{\mathcal{L}}(\Omega)+ \overline{\mathcal{L}}(-\Omega)\big)\hat{\sigma}_z,\nonumber
\end{align}
where
\begin{align}
    \overline{\mathcal{L}}(\Omega) &= \frac{1}{(2\pi)^3}\!\int \dd |\bm k| \, |\bm k|^\gamma |\tilde{f}(|\bm k|)|^2 |\tilde{\chi}(|\bm k| + \Omega)|^2,\nonumber\\
    \overline{\mathcal{K}}(\Omega) &= \frac{1}{(2\pi)^3}\!\int \dd |\bm k| \, |\bm k|^\gamma |\tilde{f}(|\bm k|)|^2 \tilde{\chi}^*(|\bm k| - \Omega)\tilde{\chi}(|\bm k| + \Omega),\nonumber\\
    \overline{\mathcal{M}}(\Omega) &= \frac{1}{(2\pi)^3}\!\int \dd |\bm k| \, |\bm k|^\gamma |\tilde{f}(|\bm k|)|^2 Q(|\bm k|,\Omega),\label{eq:MLK}
\end{align}
with $Q(|\bm k|,\Omega)$ is defined in Eq.~\eqref{eq:Qkomega}, and $\gamma = 1$ for the amplitude coupled UDW model and $\gamma = 3$ for the derivative coupling. Notice that in the derivative coupling case, $\overline{\mathcal{L}}(\Omega) ={\mathcal{L}}(\Omega)$ and $\overline{\mathcal{M}}(\Omega) = {\mathcal{M}}(\Omega)$. We can start the comparison with the spin-magnetic model by noticing that for the amplitude coupling the terms in Eqs.~\eqref{eq:MLK} have two fewer powers of $|\bm k|$ in the integrand when compared to the spin coupling (Eqs.~\eqref{eq:Lomega} and~\eqref{eq:Momega}). Again, this difference is due to the different units in the field and smearing functions and it is not present in the derivative coupling. We also notice that the prefactor on the first order corrections in the UDW model is $\pi$ rather than $4\pi/3$, as we see in Eq.~\eqref{eq:rhoGapSpin} for the spin model. This difference is due to the sum over polarizations in the electromagnetic case, which give extra contributions.

Apart from the prefactor and the powers of $|\bm k|$, comparing Eqs.~\eqref{eq:rhoDUDW} and~\eqref{eq:rhoGapSpin}, we see that the $z$ component of the Bloch vector of the UDW qubit and the spin model behave exactly in the same way. This is supporting evidence of the traditional claim that the UDW model captures the essence of absorption and emission in quantum field theories~\cite{eduardoOld2013}. The dependence of the final state on $\overline{\mathcal{M}}(\Omega)$ and $\mathcal{M}(\Omega)$ is also the same in both models. On the other hand, the components $x$ and $y$ (orthogonal to the direction determined by the free Hamiltonian) behave differently in the UDW case, as their dynamics involve $\overline{\mathcal{K}}(\Omega)$, which have no analogue in the spin magnetic field detector model. This term is also responsible for breaking the time translation symmetry $\chi(t)\mapsto \chi(t-t_0)$, under which $\overline{\mathcal{K}}(\Omega)\mapsto e^{2 \ii \Omega t_0}\overline{\mathcal{K}}(\Omega)$. Also notice that the noise terms $\mathcal{L}(0)$ of Eq.~\eqref{eq:rhoGapSpin} in the $x$ and $y$ components are not present in the final state of the UDW model.

In many ways the local interaction of a gapped spin with an external magnetic field is simpler than the UDW interaction. Not only does it require fewer parameters to fully characterize the leading order final state of the system, it also possesses more symmetries such as time translation symmetry\footnote{Notice that the UDW model is time-translation invariant for the ground excited components, or when $\Omega = 0$. The reason why other components do not technically possess this symmetry is the fact that if one starts in an initial state that is not an eigenstate of the free Hamiltonian, the free dynamics will rotate it, changing its alignment with respect to the monopole $\hat{\sigma}_x(t_0)$.}. The spin-magnetic field coupling is also spherically symmetric in the degenerate case, in the sense that given a group element $g\in \text{SU}(2)$ that implements rotations in the rest space of $u^\mu$, the time evolution operator $\hat{U}_I$ is invariant under $g$. This is not the case in the two-level UDW model, as the coupling privileges a direction in the Bloch sphere\footnote{for other versions of UDW model (for example the Harmonic oscillator UDW detector) of course one can choose the UDW coupling to also be $\text{SO}(3)$-symmetric, but this is not possible for qubit detectors.}. 

Finally, it is worth highlighting the explicit differences between the models in the case $\Omega = 0$, where the fact that the two-level UDW coupling privileges a direction in the Bloch sphere can be easily seen. The leading order final state in Eq.~\eqref{eq:rhoDUDW} becomes
\begin{align}
    &\hat{\rho}_\tc{d} = \hat{\rho}_{\tc{d},0}- 2\pi \lambda^2\overline{\mathcal{L}}(0)  \left(a^y \hat{\sigma}_y+ a^z \hat{\sigma}_z\right) \label{eq:rhoDUDWgapless}
\end{align}
when one sets $\Omega = 0$. To obtain Eq.~\eqref{eq:rhoDUDWgapless}, we used that $\overline{\mathcal{K}}(0) = \overline{\mathcal{M}}(0) = \overline{\mathcal{L}}(0)\in\mathbb{R}$. In the gapless case, the $x$ component of an UDW detector in the Bloch sphere remains unchanged. One can see it at leading order in Eq.~\eqref{eq:rhoDUDWgapless}, but it is also true non-perturbatively.

Indeed, when $\Omega = 0$, one can compute the non-perturbative dynamics of a UDW detector using the Magnus expansion~\cite{magnus,magnusReview} (see, among others,~\cite{Landulfo} for its application to the UDW model). The non-perturbative computation relies on the fact that the UDW Hamiltonian density~\eqref{eq:hiUDW} satisfies 
\begin{equation}
    [\hat{h}_I(\mf x), \hat{h}_I(\mf x')] = \lambda^2 \Lambda(\mf x) \Lambda(\mf x')[\hat{\phi}(\mf x),\hat{\phi}(\mf x')],
\end{equation}
and $[\hat{\phi}(\mf x),\hat{\phi}(\mf x')]\propto \openone$ implies
\begin{equation}\label{eq:tripleComm}
     [[\hat{h}_I(\mf x), \hat{h}_I(\mf x')],\hat{h}_I(\mf x'')] = 0,
\end{equation} 
which cancels all orders higher than second in the Magnus expansion. On the other hand, the Hamiltonian density for the interaction of a spin with electromagnetism~\eqref{eq:hIspinEMdetector} does not fulfill Eq.~\eqref{eq:tripleComm}. Indeed, the Hamiltonian density of Eq.~\eqref{eq:hIspinEMdetector} satisfies
\begin{align}
    [\hat{h}_I(\mf x), \hat{h}_I(\mf x')] = &\left(\frac{q}{2\pi}\right)^2 \Lambda(\mf x) \Lambda(\mf x') \delta^{ij} [\hat{B}_i(\mf x), \hat{B}_j(\mf x')]\nonumber\\ +& \frac{1}{2} \left(\frac{q}{2\pi}\right)^2 \Lambda(\mf x) \Lambda(\mf x')[\hat{\sigma}^i,\hat{\sigma}^j]\{\hat{B}_i(\mf x), \hat{B}_j(\mf x')\}.
\end{align}
The second term above (which is not present for a two-level UDW model) comes from the fact that the operator $\Lambda(\mf x) \hat{\bm \sigma}$ does not generally commute with itself at all spacelike separated events $\mf x$ and $\mf x'$. This is not an exclusive feature of the spin-magnetic coupling: finite-sized particle detector models usually introduce detector operators which do not fulfill the microcausality condition~\cite{us,us2,PipoFTL,mariaPipoNew}. This is a natural consequence of the non-relativistic approximations for the internal degrees of freedom of the detector that lead to effective detector models. However, in the two-level UDW model, this only happens when the detector has non-trivial internal dynamics ($\Omega\neq 0$). In the spin-magnetic model, the dependence on the anti-commutator $\{\hat{B}_i(\mf x), \hat{B}_j(\mf x')\}$ makes it so that Eq.~\eqref{eq:tripleComm} does not hold, implying that the Magnus expansion does not terminate, and cannot easily be used to obtain non-perturbative results in this case.

In summary, the spin magnetic coupling model is actually simpler to work with than the two-level UDW model when the detector has a finite gap. Conversely, the two-level UDW model is admittedly simpler than the spin magnetic coupling model in the case of degenerate detectors.

\section{Conclusions}\label{sec:Conclusions}

We have shown how a relativistic quantum field theory describing an electron in an $s$ orbital under a central potential can be reduced to an effective spin system coupled to the quantum degrees of freedom of the magnetic field. Namely we showed how to obtain the interaction of a spin with the magnetic field as a smeared Zeeman Hamiltonian proportional to $\hat{\bm \sigma} \cdot \hat{\bm B}(\mf x)$, starting from a relativistic QFT formulation. Within this framework, the spin degrees of freedom are described as excitations of a relativistic fermionic QFT, identifying an effective $\mathfrak{su}(2)$ subalgebra in the QFT formulation that corresponds to a relativistic description of the electron spin observables. 

This interaction can be compared with the `particle detector models' often employed in the literature on relativistic quantum information and QFT to model, for example, the electric dipole coupling of atoms to the electromagnetic field~\cite{richard}. In particular, we compared the spin-magnetic coupling with the conventional two-level Unruh-DeWitt particle detector. We found that when an external classical magnetic field induces a splitting of spin energy levels, the spin components aligned with $\bm B$ are well described by the UDW model, while the perpendicular components exhibit distinct behavior. This difference can be traced back to the fact that the UDW model's coupling with a scalar field is unable to isotropically couple to qubit observables in the Bloch sphere. As a consequence, if the two spin states have a non-zero energy gap, the spin detector model is a more symmetric, and in many ways simpler, version of the UDW model, yielding leading order results that are characterized by less parameters. On the other hand, the gapless spin model is more complex than the gapless UDW model, due to the fact that its interaction Hamiltonian does not commute with itself at different spacelike separated points. This prevents some of the usual non-perturbative methods from being applied to the spin interaction with a magnetic field.

Describing the coupling of a spin with the magnetic field from a fundamental QFT formulation---rather than prescribing it as an effective model in non-relativistic physics---opens the doors to implementations of relativistic quantum information protocols with electron spins. 
On the theoretical side, this paves the way to connecting operational~\cite{chicken} and formal~\cite{FewsterVerch,fewster2} perspectives of measurements in quantum field theory. On the experimental side, it provides a platform to connect relativistic quantum information protocols, such as entanglement harvesting and quantum energy teleportation~\cite{teleportation,teleportation2014,nichoTeleport,teleportExperiment}  to experimental testbeds in atomic physics involving spin-magnetic interactions. It would also be interesting to analyze the response of accelerated detectors with this model, as the manifestation of the Unruh effect and thermal phenomena present slight subtleties when the detectors are fermionic in nature~\cite{takagi,unruhfermionictobecommented,unruhcomment,unruhreplytocomment,Friis,Becattini_2013,Friis_2016,Becattini_2018,Friis2,phasetransitionunruh}.



\acknowledgements

RS thanks the PSI program for facilitating this research and contributions from Canada First Research Excellence Fund, Canada Foundation for Innovation, and Mike \& Ophelia Lazaridis. TRP acknowledges support from the Natural Sciences and Engineering Research Council of Canada (NSERC) via the Vanier Canada Graduate Scholarship. Research at Perimeter Institute is supported in part by the Government of Canada through the Department of Innovation, Science and Industry Canada and by the Province of Ontario through the Ministry of Colleges and Universities. Perimeter Institute and the University of Waterloo are situated on the Haldimand Tract, land that was promised to the Haudenosaunee of the Six Nations of the Grand River, and is within the territory of the Neutral, Anishinaabe, and Haudenosaunee people.

\onecolumngrid

\appendix

\section{Corrections to the Zeeman interaction to leading order in the fine structure constant}\label{app:derivation}

The goal of this appendix is to obtain the result of Eq.~\eqref{hflatzeemanexact}, which shows corrections to the Zeeman effect that depend on the shape of the radial function of the electron field modes corresponding to the electron state. This amounts to computing the integral in space of the function $\upphi(r)$, defined in Eq.~\eqref{eq:phir}.

In Eq.~\eqref{psi-sol-atom} we showed the form of the $s$ orbital modes of an electron in terms of the radial functions $f(r)$ and $g(r)$. When $j=1/2$ and $p=+1$ these radial functions satisfy the differential equations~\cite{RQMgreiner}
\begin{align}
    \dv{g}{r} - (E-V(r)+m_e)f(r) &= 0,\\
    \dv{f}{r} +\frac{2}{r} f(r) + (E-V(r)-m_e)g(r) &= 0,
\end{align}
where $E$ is the energy level of the given orbital. The differential equations above are valid for any central potential $V(r)$. We can obtain an expression for $\upphi(r)$ by multiplying the first equation by $g(r)$ and the second equation by $f(r)$, yielding
\begin{align}
    \frac{1}{2}\dv{g^2}{r} - (E-V(r)+m_e)f(r)g(r) &= 0,\\
    \frac{1}{2}\dv{f^2}{r} +\frac{2}{r} f(r)g(r) + (E-V(r)-m_e)f(r)g(r) &= 0.
\end{align}
Adding the two equations we find
\begin{align}
    2m_e f(r) g(r) = \frac{1}{2}\dv{}{r}\left(f^2(r) + g^2(r)\right)+ \frac{2}{r} f^2(r).
\end{align}
Integrating the result above from $\infty$ to $r$ we find
\begin{equation}
    \upphi(r) = \int_{\infty}^r \dd r' f(r') g(r') = \frac{1}{4m_e}(f^2(r) + g^2(r)) + \frac{1}{m_e} \int_\infty^r \dd r' \frac{f^2(r')}{r'},
\end{equation}
where we directly integrated the total derivative using that $f(\infty) = g(\infty) = 0$. 

The integral of $\upphi(r)$ in space is then given by
\begin{equation}
    \int \dd^3 \bm x \, \upphi(|\bm x|) = \frac{4\pi}{4m_e} \int_0^\infty \dd r \, r^2 (f^2(r) + g^2(r)) - \frac{4\pi}{m_e}\int_0^\infty \dd r  \int_r^\infty  \dd r'\frac{r^2}{r'} f^2(r'),
\end{equation}
where we picked up factors of $4\pi$ due to integration over the angular coordinates and spherical symmetry of the functions involved. Normalization of the spinor spherical harmonics and of the mode functions $\psi_{\bm N}(\bm x)$ implies that
\begin{equation}\label{eq:normfg}
    \int_0^\infty \dd r \, r^2 (f^2(r) + g^2(r)) = 1.
\end{equation}
To handle the double integral, we can reparametrize the region defined by $0<r<\infty$ and $r<r'<\infty$ as $0<r'<\infty$ and $0<r<r'$. This gives
\begin{equation}\label{eq:changerrp}
    \int_0^\infty \dd r  \int_r^\infty  \dd r'\frac{r^2}{r'} f^2(r') = \int_0^\infty \dd r'  \int_0^{r'}  \dd r\frac{r^2}{r'} f^2(r') = \frac{1}{3}\int_0^\infty \dd r' (r')^2 f^2(r') = \frac{1}{3}\int_0^\infty \dd r\, r^2 f^2(r).
\end{equation}
Combining the results of Eqs.~\eqref{eq:normfg} and~\eqref{eq:changerrp}, we find
\begin{equation}
    \int \dd^3 \bm x \, \upphi(|\bm x|) = \frac{\pi}{m_e} - \frac{4\pi}{3m_e}\int_0^\infty \dd r\, r^2 f^2(r).
\end{equation}
Thus, with a constant magnetic field we find
\begin{equation}
    \hat{H}_I(t) = - \frac{q}{2\pi} \int d^3\bm x\, \upphi(r)\, \hat{\bm{\sigma}}\cdot\bm{B}(t) = - \frac{q}{2m_e}\left(1 - \frac{4}{3}\int_0^\infty \dd r\, r^2 f^2(r)\right)\hat{\bm{\sigma}}\cdot\bm{B}(t).
\end{equation}

Importantly, this result holds true for modes with $j=1/2$ and $p=+1$ defined by any central potential $V(r)$. The corrections are then entirely determined by the function $f(r)$, and are of the order of the inverse of the product of the mass of the electron and the effective localization of the bound states.

\section{The Dynamics of a spin coupled to electromagnetism with internal dynamics.}\label{app:the full dynamics}

In this appendix we compute the leading order correction to the state of a spin system interacting with the vacuum of the quantum magnetic field when there is an energy an energy gap $\Omega$ between two spin states, as discussed in the setup of Subsection~\ref{sub:gapped}. The leading order corrections to the state will be given by
\begin{equation}
    \delta\hat{\rho}_\tc{d} = \tr_B\left( \hat{U}_I^{(1)} \hat{\rho}_{0}\hat{U}_I^{(1)\dagger}\right) + \tr_B\left( \hat{U}_I^{(2)} \hat{\rho}_{0}\right) + \tr_B\left(\hat{\rho}_{0}\hat{U}_I^{(2)\dagger}\right),
\end{equation}
where
\begin{align}
    \tr_B\left( \hat{U}_I^{(1)} \hat{\rho}_{0}\hat{U}_I^{(1)\dagger}\right) \label{eq:U1rhoU1t}&= \left(\frac{q}{2\pi}\right)^{\!2}\int \dd V \dd V' \Lambda(\mf x) \Lambda(\mf x') \hat{\sigma}_j(t')\hat{\rho}_{\tc{d},0} \hat{\sigma}_i(t) \langle \hat{B}^i(\mf x)\hat{B}^j(\mf x') \rangle,\\
    \tr_B\left( \hat{U}_I^{(2)} \hat{\rho}_{0}\right)\label{eq:U2rhot}&= \!-\left(\frac{q}{2\pi}\right)^{\!2}\!\!\!\int\! \!\dd V \dd V' \Lambda(\mf x) \Lambda(\mf x') \hat{\sigma}_i(t) \hat{\sigma}_j(t') \hat{\rho}_{\tc{d},0} \langle \hat{B}^i(\mf x) \hat{B}^j(\mf x')\rangle \theta(t-t'),\\
    \tr_B\left(\hat{\rho}_{0}\hat{U}_I^{(2)\dagger}\right) &= \label{eq:rhoU2t}\!-\left(\frac{q}{2\pi}\right)^{\!2}\!\!\!\int\! \!\dd V \dd V' \Lambda(\mf x) \Lambda(\mf x')  \hat{\rho}_{\tc{d},0}\hat{\sigma}_i(t) \hat{\sigma}_j(t') \langle \hat{B}^i(\mf x) \hat{B}^j(\mf x')\rangle \theta(t'-t).
\end{align}

Using the expression for the two-point function of the magnetic field in Eq.~\eqref{eq:Bpropag}, and writing the spacetime smearing function as $\Lambda(\mf x) = \chi(t)\upphi(\bm x)$ (Eq.~\eqref{eq:spacetimeSmearingAtom}), we can write the leading order correction as
\begin{equation}
    \delta \hat{\rho}_\tc{d} = \left(\frac{q}{2\pi}\right)^2 \int \dd^3 \bm x \dd^3 \bm x' \dd t \dd t' \chi(t) \chi(t') \upphi(\bm x)\upphi(\bm x') \frac{1}{(2\pi)^3}\!\!\int \frac{\dd^3 \bm k}{2|\bm k|} |\bm k|^2 e^{- \ii |\bm k| (t-t')}e^{\ii\bm k\cdot  (\bm x-\bm x')} \sum_{s=1}^2\epsilon^i_s(\bm k) \epsilon_s^j(\bm k)\hat{R}_{ij}(t,t'),
\end{equation}
where
\begin{equation}
    \hat{R}_{ij}(t,t') = \hat{\sigma}_j(t') \hat{\rho}_{\tc{d},0}\hat{\sigma}_i(t) -\hat{\sigma}_i(t) \hat{\sigma}_j(t') \hat{\rho}_{\tc{d},0}\theta(t-t')-\hat{\rho}_{\tc{d},0} \hat{\sigma}_i(t) \hat{\sigma}_j(t')\theta(t'-t).
\end{equation}
By noticing that 
\begin{align}
    \int \dd^3\bm x \dd^3\bm x' \upphi(\bm x)\upphi(\bm x') e^{\ii \bm k\cdot (\bm x - \bm x')} = |\tilde{\upphi}(|\bm k|)|^2,
\end{align}
and performing the integral over $\bm k$ in spherical coordinates $(|\bm k|,\vartheta,\varphi)$, we can further rewrite the leading order correction to the state as
\begin{equation}
    \delta\hat{\rho}_\tc{d} = \left(\frac{q}{2\pi}\right)^2  \frac{1}{2} \frac{1}{(2\pi)^3}\int \dd |\bm k|\,|\bm k|^3 |\tilde{\upphi}(|\bm k|)|^2  \int \dd t \dd t' \chi(t) \chi(t')  e^{- \ii |\bm k| (t-t')} \int \dd \vartheta \dd \varphi \sin\vartheta\sum_{s=1}^2\epsilon^i_s(\bm k) \epsilon_s^j(\bm k)\hat{R}_{ij}(t,t').\label{eq:deltarhod}
\end{equation}

We now focus on the integral over the angular variables $\vartheta$ and $\varphi$. Writing
\begin{align}
    \hat{\rho}_{\tc{d},0} = \frac{1}{2}\begin{pmatrix}
        1 + a^z & a^x - \ii a^y \\
        a^x + \ii a^y & 1 - a_z
    \end{pmatrix},
\end{align}
we find
\begin{align}
    \int \dd \vartheta \dd \varphi \sin\vartheta \epsilon_1^{i}(\bm k)\epsilon_1^{j}(\bm k) \hat{\sigma}_i(t) \hat{\sigma}_j(t') \hat{\rho}_{\tc{d},0}\theta(t-t') &= \pi \begin{pmatrix}
        (1+a^z)e^{\ii\Omega(t - t')} & (a^x - \ii a^y)e^{\ii\Omega(t - t')}\\
        (a^x + \ii a^y)e^{-\ii\Omega(t - t')} & (1-a^z)e^{-\ii\Omega(t - t')}
    \end{pmatrix}\theta(t-t'),\label{eq:mathematica}\\
    \int \dd \vartheta \dd \varphi \sin\vartheta \epsilon_1^{i}(\bm k)\epsilon_1^{j}(\bm k) \hat{\rho}_{\tc{d},0} \hat{\sigma}_i(t) \hat{\sigma}_j(t')\theta(t'-t) &= \pi \begin{pmatrix}
        (1+a^z)e^{\ii\Omega(t - t')} & (a^x - \ii a^y)e^{-\ii\Omega(t - t')}\\
        (a^x + \ii a^y)e^{\ii\Omega(t - t')} & (1-a^z)e^{-\ii\Omega(t - t')}
    \end{pmatrix}\theta(t'-t),\nonumber\\
    \int \dd \vartheta \dd \varphi \sin\vartheta \epsilon_1^{i}(\bm k)\epsilon_1^{j}(\bm k) \hat{\sigma}_j(t') \hat{\rho}_{\tc{d},0}\hat{\sigma}_i(t) &= -2\pi \begin{pmatrix}
        (1-a^z)e^{-\ii\Omega(t - t')} & 0\\
        0 & (1+a^z)e^{\ii\Omega(t - t')}
    \end{pmatrix},\nonumber\\
    \int \dd \vartheta \dd \varphi \sin\vartheta \epsilon_2^{i}(\bm k)\epsilon_2^{j}(\bm k) \hat{\sigma}_i(t) \hat{\sigma}_j(t') \hat{\rho}_{\tc{d},0}\theta(t-t') &= \frac{\pi}{3} \begin{pmatrix}
        (1+a^z)(2+e^{\ii\Omega(t - t')}) & (a^x - \ii a^y)(2+e^{\ii\Omega(t - t')})\\
        (a^x + \ii a^y)(2+e^{-\ii\Omega(t - t')}) & (1-a^z)(2+e^{-\ii\Omega(t - t')})
    \end{pmatrix}\theta(t-t'),\nonumber\\
    \int \dd \vartheta \dd \varphi \sin\vartheta \epsilon_2^{i}(\bm k)\epsilon_2^{j}(\bm k) \hat{\rho}_{\tc{d},0} \hat{\sigma}_i(t) \hat{\sigma}_j(t')\theta(t'-t) &= \frac{\pi}{3} \begin{pmatrix}
        (1+a^z)(2+e^{\ii\Omega(t - t')}) & (a^x - \ii a^y)(2+e^{-\ii\Omega(t - t')})\\
        (a^x + \ii a^y)(2+e^{\ii\Omega(t - t')}) & (1-a^z)(2+e^{-\ii\Omega(t - t')})
    \end{pmatrix}\theta(t'-t),\nonumber\\
    \int \dd \vartheta \dd \varphi \sin\vartheta \epsilon_2^{i}(\bm k)\epsilon_2^{j}(\bm k)  \hat{\sigma}_j(t') \hat{\rho}_{\tc{d},0}\hat{\sigma}_i(t) &= \frac{2\pi}{3} \begin{pmatrix}
        2(1+a^z)+(1-a^z)e^{-\ii\Omega(t - t')} & -2(a^x-\ii a^y)\\
        -2(a^x+\ii a^y) & 2(1-a^z)+(1+a^z)e^{\ii\Omega(t - t')}
    \end{pmatrix}.\nonumber
\end{align}
Combining the results above, we obtain
\begin{align}
    \sum_{s=1}^2\int& \dd \vartheta \dd \varphi \sin\vartheta \epsilon_s^{i}(\bm k)\epsilon_s^{j}(\bm k) \left(\hat{\sigma}_j(t') \hat{\rho}_{\tc{d},0}\hat{\sigma}_i(t) - \hat{\sigma}_i(t) \hat{\sigma}_j(t') \hat{\rho}_{\tc{d},0}\theta(t-t')-\hat{\rho}_{\tc{d},0} \hat{\sigma}_i(t) \hat{\sigma}_j(t')\theta(t'-t)\right)\\
    &= \frac{8\pi}{3}\begin{pmatrix}
        a^z(e^{\ii \Omega (t-t')} + e^{-\ii \Omega (t-t')}) + e^{\ii \Omega(t-t')}-e^{-\ii \Omega(t-t')} & -a^x(1 + e^{\ii \Omega |t-t'|}) + \ii a^y(1 + e^{\ii \Omega |t-t'|})\\
        -a^x(1 + e^{-\ii \Omega |t-t'|}) - \ii a^y(1 + e^{-\ii \Omega |t-t'|}) & -a^z(e^{\ii \Omega (t-t')} + e^{-\ii \Omega (t-t')}) - e^{-\ii \Omega(t-t')}+e^{\ii \Omega(t-t')}
    \end{pmatrix},
\end{align}
where we used $f(t-t')\theta(t-t') + f(t'-t)\theta(t'-t) = f(|t-t'|)$ and $\theta(t-t') + \theta(t'-t) = 1$ to combine the results of Eq.~\eqref{eq:mathematica} that involve the Heaviside theta function.
Now using
\begin{equation}
    \int \dd t \dd t' \chi(t) \chi(t') e^{- \ii \Omega (t-t')}e^{- \ii |\bm k|(t-t')} = |\tilde{\chi}(|\bm k|+\Omega)|^2,
\end{equation}
and defining
\begin{equation}
    Q(|\bm k|,\Omega) = \int \dd t \dd t' \chi(t)\chi(t') e^{\ii \Omega |t-t'|}e^{- \ii |\bm k|(t-t')},
\end{equation}
we find
\begin{align}
    &\int \dd t \dd t' \chi(t) \chi(t') e^{-\ii |\bm k|(t-t')}\sum_{s=1}^2\int  \dd \vartheta \dd \varphi \sin\vartheta \epsilon_s^{i}(\bm k)\epsilon_s^{j}(\bm k) \hat{R}_{ij}(t,t')\\
    &=-\frac{8\pi}{3} \begin{pmatrix}
         \scriptstyle{a^z(|\tilde{\chi}(|\bm k|-\Omega)|^2 + |\tilde{\chi}(|\bm k|+\Omega)|^2) + |\tilde{\chi}(|\bm k|-\Omega)|^2-|\tilde{\chi}(|\bm k|+\Omega)|^2} & \scriptstyle{a^x(|\tilde{\chi}(|\bm k|)|^2 + Q(|\bm k|,\Omega) - \ii a^y(|\tilde{\chi}(|\bm k|)|^2 + Q(|\bm k|,\Omega))}\\
        \scriptstyle{a^x(|\tilde{\chi}(|\bm k|)|^2 + Q(|\bm k|,-\Omega) + \ii a^y(|\tilde{\chi}(|\bm k|)|^2 + Q(|\bm k|,-\Omega))} & \scriptstyle{-a^z(|\tilde{\chi}(|\bm k|-\Omega)|^2 + |\tilde{\chi}(|\bm k|+\Omega)|^2) - |\tilde{\chi}(|\bm k|-\Omega)|^2+|\tilde{\chi}(|\bm k|+\Omega)|^2}
    \end{pmatrix}.
\end{align}
Combining the result above with Eq.~\eqref{eq:deltarhod}, and using the definitions of $\mathcal{L}(\Omega)$ and $\mathcal{M}(\Omega)$ of Eqs.~\eqref{eq:Lomega} and~\eqref{eq:Momega}, together with the fact that $\mathcal{M}(-\Omega) = \mathcal{M}(\Omega)^*$, we obtain 
\begin{equation}
    \delta \hat{\rho}_\tc{d} = - \frac{4\pi}{3}\left(\frac{q}{2\pi}\right)^2 \begin{pmatrix}
        a^z (\mathcal{L}(\Omega) + \mathcal{L}(-\Omega)) - \mathcal{L}(\Omega) + \mathcal{L}(-\Omega) & a^x(\mathcal{L}(0) + \mathcal{M}(\Omega)) - \ii a^y (\mathcal{L}(0) + \mathcal{M}(\Omega))\\
        a^x(\mathcal{L}(0) + \mathcal{M}(\Omega)^*) + \ii a^y (\mathcal{L}(0) + \mathcal{M}(\Omega)^*) &  -a^z (\mathcal{L}(\Omega) + \mathcal{L}(-\Omega)) + \mathcal{L}(\Omega) - \mathcal{L}(-\Omega)
    \end{pmatrix}.
\end{equation}
Decomposing the off-diagonal terms above into their real and imaginary parts, we can put the correction in the form presented in Eq.~\eqref{eq:rhoGapSpin}.

\twocolumngrid

\bibliography{references.bib}

\end{document}